\RequirePackage{fix-cm}
\documentclass[twocolumn,epjc3]{svjour3}  

\journalname{Eur. Phys. J. E}

\RequirePackage{latexsym}
\RequirePackage[numbers,sort&compress]{natbib}
\RequirePackage[colorlinks=true,citecolor=blue,urlcolor=blue,linkcolor=blue,urlcolor=blue]{hyperref}

\usepackage[title]{appendix}

\usepackage{graphicx}
\usepackage{color}

\usepackage{amssymb}
\usepackage{amsmath}
\usepackage{amsthm}
\usepackage{bm}

\usepackage{lscape}
\usepackage{subcaption}
\usepackage{algorithm}
\usepackage{tabularx}
\usepackage[noend]{algpseudocode}

\usepackage{ulem}
\usepackage[percent]{overpic}

\definecolor{orange}{rgb}{1,0.5,0}
\definecolor{forestgreen}{rgb}{0.13, 0.55, 0.13}
\definecolor{bittersweet}{rgb}{1.0, 0.44, 0.37}
\definecolor{chartreuse}{rgb}{0.87, 1.0, 0.0}
\definecolor{darkorchid}{rgb}{0.6, 0.2, 0.8}

\newcommand{\Sg}{\mathcal{S}}

\def\mR{\mathbb{R}}

\begin{document}

\makeatletter
\@fleqnfalse
\@mathmargin\@centering
\makeatother

\title{Towards learning Lattice Boltzmann collision operators}

\author{ Alessandro Corbetta \thanksref{inst1}     \and
         Alessandro Gabbana  \thanksref{inst1, e1} \and
         Vitaliy Gyrya       \thanksref{inst2}     \and
         Daniel Livescu      \thanksref{inst2}     \and
         Joost Prins         \thanksref{inst1}     \and
         Federico Toschi     \thanksref{inst1,inst3}   
       } 

\thankstext{e1}{e-mail: a.gabbana@tue.nl}

\institute{Eindhoven University of Technology, 5600 MB Eindhoven,~ The Netherlands \label{inst1}
           \and
           Los Alamos National Laboratory, NM 87545 Los Alamos, ~ USA \label{inst2}
           \and Consiglio Nazionale della Ricerche-IAC, Rome, Italy \label{inst3}
}

\date{Received: date / Accepted: date}

\abstractdc{
          In this work we explore the possibility of learning from data collision operators 
          for the Lattice Boltzmann Method using a deep learning approach.
          We compare a hierarchy of designs of the neural network (NN) collision operator 
          and evaluate the performance of the resulting LBM method in reproducing time dynamics of several canonical flows.
          In the current study, as a first attempt to address the learning problem, the data was generated by a single relaxation time BGK operator.           
          We demonstrate that vanilla NN architecture has very limited accuracy.
          On the other hand, by embedding physical properties, such as conservation laws and symmetries, 
          it is possible to dramatically increase the accuracy by several orders of magnitude and 
          correctly reproduce the short and long time dynamics of standard fluid flows.
}

\maketitle

\section{Introduction}\label{sec:intro}

The Lattice Boltzmann Method (LBM) is a computationally efficient method for 
the simulation of fluid flows in a wide range of regimes.
LBM allows solving a set of macroscopic equations via the time evolution of
a (minimal) discrete version of the continuum Boltzmann equation, 
following the stream and collide paradigm.

While its original formulation targets mostly isothermal 
weakly compressible fluid flows,
over the years several algorithmic developments have allowed 
extending the method to support the simulation of a wide range of complex flows,
such as 
multi-phase~\cite{shan-pre-1993,sbragaglia-pre-2007},
turbulence~\cite{chen-science-2003}, 
thermo-hydrodynamics~\cite{philippi-pre-2006,scagliarini-pof-2010}, 
non-Newtonian flows~\cite{aharonov-grl-1993,gabbanelli-pre-2005},
radiative transport~\cite{asinari-nhtb-2010},
semi-classical fluids~\cite{coelho-cf-2018}, 
relativistic flows~\cite{gabbana-pr-2020},
and many others~\cite{succi-book-2018}. 
Most of these algorithmic enhancements have targeted the modeling of the collision process
and, as a result, a large variety of collision models have been proposed to extend the 
applicability and overcome the shortcomings of the standard LBM.
Notable examples extending the single relaxation time Bhatnagar-Gross-Krook (BGK) collision operator~\cite{bhatnagar-pr-1954}
are given by the two relaxation times (TRT)~\cite{ginzburg-ccp-2008}, 
multi-relaxation time (MRT) ~\cite{dumieres-paa-1992, lallemand-pre-2000},
which can be combined with regularization procedures~\cite{latt-mcs-2006, zhang-pre-2006, mattila-pof-2017}, 
and local viscous corrections, 
ensuring the validity of the H-theorem after the velocity discretization~\cite{karlin-prl-1998, ansumali-pre-2002}. 
More recent developments have taken into consideration 
the ellipsoidal statistical BGK ~\cite{meng-jfm-2013} 
and the Shakov model~\cite{ambrus-pre-2018}, 
which allow to decouple the thermal relaxation from the viscous one. 
They also made possible to compute equilibrium distributions numerically, 
in principle, allowing to reproduce an arbitrary number of moments of the Maxwell-Boltzmann distribution~\cite{latt-ptrsa-2020}.
For a comprehensive review comparing collision models for LBM
the interested reader is referred to~\cite{coreixas-pre-2019}.

In recent years, there has been an increased interest in adoption of machine learning (ML) models, typically, of artificial neural networks (NN), 
to approximate various kernels/operators in the simulation of physical systems. 
Artificial neural networks form a class of nonlinear parametric models satisfying universal approximation property~\cite{hornik-nn-1989}. 
This property coupled with
efficient computational tools for automatic differentiation and sensitivity analysis of forward and backward propagation,
in the last decade,
has led to outstanding results in such fields as computer vision \cite{sebe2005machine} and natural language processing \cite{wieting2016ICLR}. 

However, until recently,
the biggest achievements of ML in scientific environment have been limited
to approaches that are data-driven but agnostic to traditional scientific modeling of the underlying physics. 
Integrating the modern ML with physical modeling is the major challenge of what we call today Physics-Informed Machine Learning (PIML)~\cite{2018PIML-LANL,Karniadakis2021Physic-Informed}. 
In particular, in fluid dynamics, there has been significant PIML activity in recent years. 
Examples include embedding physical constraints, such as Galilean invariance and rotational invariance, into the closure model \cite{ling2016reynolds,tian2021physics} and 
PIML models infusing physical constraints into the neural networks \cite{wang2017physics,mohan2020div}. 
Other efforts on turbulence modeling are summarized in \cite{duraisamy2019turbulence,ortali2022numerical}. 
In addition to developing closure models, 
novel ML approaches have been used to learn turbulence dynamics \cite{mohan2020jot}, 
where a Convolutional Long Short Term Memory (ConvLSTM) Neural Network was developed to learn spatial-temporal turbulence dynamics; 
study super-resolution allowing to reconstruct turbulence fields using under-resolved data \cite{fukami2019super}; 
use Neural Ordinary Differential Equation (Neural ODE) for turbulence forecasting \cite{portwood2019turbulence};
or measure~\cite{corbetta2021deep}, model and control flows~\cite{beintema2020controlling}.

Up to now, very few works have proposed applications of ML to LBM. 
Most of these have been focusing on accelerating the calculation of steady-state flows
using convolutional neural networks~\cite{hennigh-arxiv-2017, guo-proc-2016, wang-tpm-2021},
while Bedrunka et al.~\cite{bedrunka-hpc-2021} employed a
fully connected feed-forward neural network to tune the parameters 
of a MRT collision operator.

Since LBM entails a mesoscopic representation,
it employs substantially more degrees of freedom 
(i.e. the number of discrete particle distribution functions)
than the macroscopic observables of interest. 
These extra degrees of freedom suggest a possibility of using ML
to encode more information in the model in order, for example, 
to extend its applicability, accuracy, and enhance the numerical stability. 

In this work we take a first step in this direction, 
and consider the problem of learning a custom collision operator from reference data.
The collision operator will be represented by a NN
that takes as inputs pre-collision and return post-collision populations. 
As a proof-of-concept we evaluate different neural network architectures 
to identify design choices that improve performance of the learned collision operator.
To make performance evaluation more straightforward
we consider a large synthetic dataset containing 
pre- and post-collision populations pairs
that itself was generated by a collision operator, 
specifically the BGK collision operator.
%
In theory, in the limit of infinite data and infinite training resources 
it should be possible to recover the underlying operator.
On the other hand, in practice, there will always be an error that 
(as we show later) significantly depends on the architecture of the NN.
We show that constraining the NN to respect physics properties 
such as conservation laws and symmetries is key for accuracy.
We evaluate the accuracy of the learned collision operator 
on both single-step (static) collision, as well as
multi-step (dynamic) collisions, interleaved with advection steps,
for the simulation of standard benchmarks.
The focus of this work is on exposing the main 
ingredients needed to accurately learn a collision operator from data, 
while, for the moment, no attention is paid to computational efficiency.

This article is structured as follows: 
in Section~\ref{sec:lbm}, we provide a brief description of the Lattice Boltzmann Method.
In Section~\ref{sec:ml}, we define a PIML approach for learning a collision operator
from data, focusing in particular on the embedding of relevant physical properties.
In Section~\ref{sec:results}, we report simulations results for two numerical benchmarks 
where we have replaced the collision term in LBM simulations with a neural network. Here, we
also compare the accuracy achieved by different neural network architectures.
Concluding remarks and future directions are summarized in Section~\ref{sec:conclusions}.

\section{Lattice Boltzmann Method}\label{sec:lbm}

In this section, we give a short introduction to the Lattice Boltzmann Method (LBM);
the interested reader is referred to, e.g., Ref.~\cite{succi-book-2018,kruger-book-2017} 
for a thorough introduction.

LBM simulates the evolution of macroscopic quantities (such as density
and velocity) 
through a mesoscopic approach based on the synthetic
dynamics of a set of discrete velocity distribution functions 
\[f_i(\bm{x},t), \ i = 0, \dots, q-1,\] 
to which we will refer
as lattice populations.

At each grid node $\bm{x}$, the lattice populations are defined 
along the discrete components of the stencil $\{ \xi_i \}, \ i = 1, \dots, q-1$.
It is customary to distinguish between different LBM schemes using the
D$d$Q$q$ nomenclature, in which $d$ refers to the number of spatial
dimensions and $q$ to the number of discrete components.
%

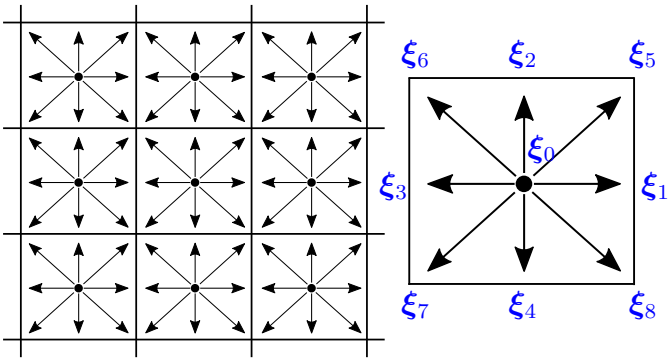
\begin{figure}[htb]
  \centering
  \begin{overpic}[width =.99\columnwidth]{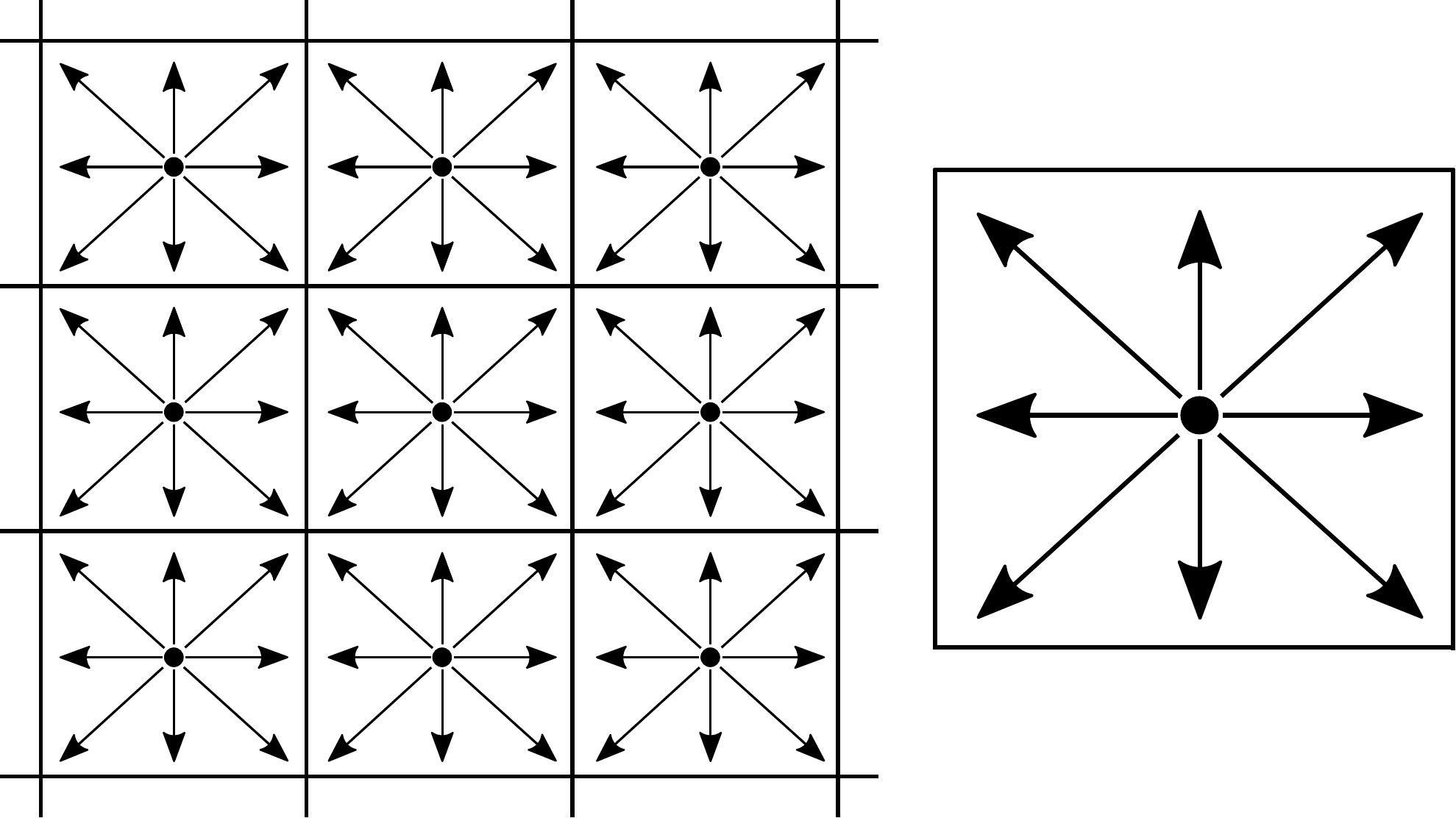}
    \put( 83, 32){\color{blue} \large $\bm{\xi}_0$}
    \put(101, 26){\color{blue} \large $\bm{\xi}_1$}
    \put( 80, 47){\color{blue} \large $\bm{\xi}_2$}
    \put(59.5,26){\color{blue} \large $\bm{\xi}_3$}
    \put( 80,  7){\color{blue} \large $\bm{\xi}_4$}
    \put( 99, 47){\color{blue} \large $\bm{\xi}_5$}
    \put( 63, 47){\color{blue} \large $\bm{\xi}_6$}
    \put( 63,  7){\color{blue} \large $\bm{\xi}_7$}
    \put( 99,  7){\color{blue} \large $\bm{\xi}_8$}       
  \end{overpic}
  \caption{Example of a 3x3 LBM grid (with a single grid point shown on 
           the right hand side) making use of the D2Q9 model
           where the lattice populations can move along 9 possible directions.
          }\label{fig:d2q9}
\end{figure}

In this work we adopt the D2Q9 model, based on the stencil in Fig.~\ref{fig:d2q9},
where populations can move along 9 possible directions, defined by the following
discrete velocity vectors:
\begin{equation*}
    \bm{\xi}_i= 
    \begin{cases}
        (0,0) & i = 0,\\
        (1,0),( 0,1),(-1, 0),(0,-1) & i = 1,2,3,4,\\
        (1,1),(-1,1),(-1,-1),(1,-1) & i = 5,6,7,8.
    \end{cases}
\end{equation*}
In general, the velocity sets, $\bm{\xi}_{i}$, are chosen such that 
any spatial vector  $\bm{\xi}_{i} \Delta t$ points from one lattice
site to a neighboring lattice site. 
This guarantees that the populations $f_i$ always reach another
lattice site during a time step $\Delta t$. 

The time evolution of each lattice population is ruled by the 
lattice Boltzmann equation which, in the absence of external forces, reads as:
\begin{equation}\label{eq:lbe}
  f_i(\bm{x}+ \bm{\xi}_{i} \Delta t , t + \Delta t) 
  -
  f_i(\bm{x},t)
  = 
  \Omega\left( f_i(\bm{x},t) \right),
\end{equation}
where $\Omega$ is the collision operator. 
Among various possible choices, in this work we 
consider the BGK~\cite{bhatnagar-pr-1954} operator
\begin{equation}\label{eq:lbgk}
  \Omega(f_i(\bm{x},t)) 
  = 
  -\frac{\Delta t}{\tau} \left( f_i(\bm{x},t) - {f}_{i}^{\rm{eq}} (\bm{x},t)\right),
\end{equation}
which models collisions as a linear relaxation process 
of the distribution function towards its equilibrium. 
Here, $\tau$ is the relaxation time, $\Delta t$ the time step,
and $f_i^{\rm eq}(\bm{x},t)$ is the discrete equilibrium distribution,
for which we employ a second order Hermite-expansion of the Maxwell-Boltzmann distribution:
\begin{equation}\label{eq:feq}
        f^{\rm{eq}}_{i}(\rho, \bm{u})   = \,  w_i \rho 
        \left( 
               1 + \frac{ \bm{u} \cdot \bm{\xi}_{i}}{c_s^2} 
                 + \frac{(\bm{u} \cdot \bm{\xi}_{i})^2 - (c_s |\bm{u}|)^2 }{2 c_s^4}
       \right),  
\end{equation}
with $w_i$ a lattice-dependent set of weighting factors. 
For the D2Q9 
\[w_0 = 4/9,\ \  w_1=w_2=w_3=w_4=1/9,\]\[ w_5=w_6=w_7=w_8=1/36.\]

In lattice units, $\Delta t=1$, while the speed of sound in the lattice for the
D2Q9 model is $c_s = 1 / \sqrt{3}$. 
Finally, $\rho$ and $\bm{u}$ indicate, respectively, 
the macroscopic density and the velocity fields. 
These macroscopic observable can be computed in terms of 
the moments of the velocity distribution functions as
\begin{align}\label{eq:moments}   
   \rho        = \sum_{i=0}^{q-1} f_i 
   %
   \qquad \text{and} \qquad
   \rho \bm{u} = \sum_{i=0}^{q-1} f_i \bm{\xi}_{i}.
\end{align}
Following an asymptotic analysis, like 
the Chapman-Enskog expansion~\cite{chapman-book-1970},
it can be shown that Eq.~\ref{eq:lbe} delivers a second order 
approximation of the Navier-Stokes equations. In particular, 
the following relation between the relaxation time parameter 
$\tau$ and the kinematic viscosity $\nu$ of the fluid holds:
\begin{equation}\label{eq:viscosity}
  \nu = \left( \tau - \frac{1}{2} \right) c_s^2  .
\end{equation}

We conclude this section by sketching the LBM algorithm.
Provided a suitable initialization of the particle distribution functions, 
each time iteration of the algorithm entails the following steps:
\begin{enumerate}
  \item Perform the streaming step:
        \begin{equation}\label{eq:streaming}
          f_i^{\rm pre}(\bm{x}, t) = f_i(\bm{x} - \bm{\xi}_{i} \Delta t , t) .
        \end{equation}  
  \item Compute the macroscopic fields using Eq.~\ref{eq:moments}
  \item Calculate the equilibrium distribution function using Eq.~\ref{eq:feq}
  \item Apply the collision operator
        \begin{align}\nonumber
          &f_i^{\rm post} 
          = f_i(\bm{x}, t + \Delta t)
          = f_i^{\rm pre}(\bm{x},t) \\
          & \quad- \frac{\Delta t}{\tau} 
          \left( f_i^{\rm pre}(\bm{x}, t) - {f}_{i}^{\rm{eq}} ( \rho(\bm{x},t), \bm{u}(\bm{x},t) ) \right) .          
          \label{eq:collide}
        \end{align}  
\end{enumerate}  

\subsection{Collision invariants and equivariances}
The operator $\Omega$ carries physical properties of the Boltzmann collision, 
which can be phrased in terms of invariances and equivariances.
Respecting these physical aspects will turn central in the performance 
of the machine learning models discussed in the next sections. 
In particular, $\Omega$ satisfies the following:
\begin{itemize}
    \item[\textbf{P1}] \textit{Scale equivariance.} 
      Scale factors $\lambda > 0$, 
      remodulating all the pre-collision populations,  
      are preserved, i.e. 
      \begin{equation}
        \Omega(\lambda f_i^{pre}) = \lambda\Omega( f_i^{pre})\ .
      \end{equation}
      In other terms, the collision is degree-1 homogeneous.
    \item[\textbf{P2}] \textit{Rotation and reflection equivariance.} 
      Generic two-dimensional collisions are equivariant with respect to 
      the 2-dimensional orthogonal group $O(2)$. This translates into the rotational 
      and mirror independence on the spectator viewpoint. 
      As we restrict to a D2Q9 lattice, this property reduces to preserving 
      the 8th order dihedral symmetry group of the lattice $D_{2n}\subset O(2)$, $n=4$. 
      This group is generated by a 90 degree rotation and a mirroring with 
      respect to symmetry axes of the cell (e.g. the $x$ axis). 
      Naming these two operations, respectively, $r$ and $s$, 
      and identifying with $I$ the identity operation, the 8 elements of $D_{8}$ are
      \begin{equation}
        D_8 = \{I,r,r^2,r^3,s,rs,r^2s,r^3s\}.
      \end{equation}
      Here, the $n$-th power indicates $n$ subsequent applications of the 
      same operator (i.e. $r^2$ is a 180 degree rotation). 
      When applied to the populations, these operators effectively 
      yields permutations of the population indices (cf. Fig.~\ref{fig:network}). 
      Finally, in formulas, rotation and mirroring equivariance of collisions reads
      \begin{equation}\label{eq:P2}
        \Omega(\sigma f_i^{pre}) = \sigma \Omega( f_i^{pre}),\ \forall \sigma \in D_8.
      \end{equation}
    \item[\textbf{P3}] \textit{Mass and momentum invariance.} 
      In the D2Q9 LBM model, mass and momentum are preserved ``exactly'' by the collision. 
      This holds thanks to the underlying Gaussian quadrature used in 
      the discretization of the velocity space~\cite{shan-pre-2010,shan-jocs-2016}:
      \begin{equation}\label{eq:conservation-laws}
        \begin{aligned}
          \sum_{i=0}^{8} \left(f_i^{\rm{post}} - f_i^{\rm{pre}} \right)            &= 0, \\
          \sum_{i=0}^{8} \left(f_i^{\rm{post}} - f_i^{\rm{pre}} \right) \bm{\xi}_i &= \bm{0}.
        \end{aligned}
      \end{equation}
\end{itemize}
Finally, we shall require \textit{positivity} (\textbf{P4}) for the post-collision
lattice populations ($f_i^{post} > 0$ for all $i$), 
since they represent discrete velocity distribution functions.

\section{Machine learning approach}\label{sec:ml}

In this section we describe a machine learning approach, hinged on a neural network, 
to approximate  the collision operator. Therefore, such a neural network will act 
as a replacement of the right hand side of Eq.~\ref{eq:lbe}.
Our learning problem aims at finding a neural network $\Omega^{\rm NN}$ 
such that $\Omega^{\rm NN}\approx \Omega$, i.e., formally,
\begin{align}
  \begin{cases}
      \tilde{f}_i^{\rm post} = \Omega^{\rm NN} (f_i^{\rm pre}), \ \ i = 0, \dots, 8,\\
      \tilde{f}_i^{\rm post}\approx f_i^{\rm post} ,
  \end{cases}
\end{align}
where the input of the network, $f_i^{\rm pre}$, is given by 
the pre-collision (post-streaming) lattice populations, 
and the network output, $\tilde{f}_i^{\rm post}$, 
targeting the post-collision populations $f_i^{\rm post}$. 

In the reminder of the section we will define:
\begin{itemize}
    \item The loss function whose minimization drives the NN training process. 
          This will also formalize our desired approximation 
          $\tilde{f}_i^{\rm post}\approx f_i^{\rm post}$.
    \item The training and testing datasets.
    \item The network architecture, addressing the strategies that 
          we considered to embed symmetries and conservations.
\end{itemize}

\paragraph{Loss function and training procedure.} 
We train the neural network to minimize 
the Mean Squared Relative Error (MSRE) between ground-truth 
post-collision populations, $ f_i^{\rm post}$, and the neural 
network approximations, $\tilde{f}_i^{\rm post}$, accumulated across the populations:
\begin{equation}\label{eq:loss}
  \rm{MSRE} 
  = 
   \sum_{i=0}^{8} 
    \left( 
            \frac{\tilde{f}_{i}^{\rm{post}} - f_{i}^{\rm{post}}}{f_{i}^{\rm{post}}} 
    \right)^2.
\end{equation}

Here, the use of a relative error metric is crucial in order to achieve good accuracy,
since in general the lattice populations take values proportional to the 
corresponding lattice weights $w_i$, and, as a consequence, an absolute error metric 
would lead to the NN learning with higher accuracy the rest-population $f_0$ 
(typically the one taking the largest value) at the expense of the others.

From an implementation perspective, we consider a mini-batch stochastic 
gradient descent approach driven by standard ADAM optimizer~\cite{kingma-arxiv-2014}.

\paragraph{Training and testing datasets.} 
In order to control the distribution of the macroscopic parameters appearing 
in the training set, we rely on synthetic data rather than actual simulation data. 
The training set consists of $N$ pairs of 9-tuples 
\begin{equation}
 \{( f_{i,k}^{\rm{pre}}, \Omega(f_{i,k}^{\rm{pre}}) ), k=1,2,\ldots,N\},   
\end{equation}
where the pre-collision distributions are generated as 
\begin{equation}
  f_i^{\rm{pre}} 
  = 
  f_i^{\rm{eq}}(\rho, \bm{u}) + f_i^{\rm{neq}} .
\end{equation} 
In the above, the equilibrium distribution $f_i^{\rm{eq}}$ is calculated using 
Eq.~\ref{eq:feq} from a set of randomly sampled macroscopic variables 
$\rho, \bm{u}$.
The non equilibrium part $f_i^{\rm{neq}}$ is such that each population is 
randomly drawn from a Gaussian distribution, after which corrections 
are introduced to ensure no contributions to lower order moments, i.e.
\begin{equation}\label{eq:fneq}
  \begin{aligned}
    \sum_{i=0}^{8} f_i^{\rm{neq}}            &= 0, \\
    \sum_{i=0}^{8} f_i^{\rm{neq}} \bm{\xi}_i &= \bm{0}.
  \end{aligned} 
\end{equation}
See Appendix~\ref{app:training-set} for further details.
\begin{table}[htb]
  \centering
  \caption{List of hyper-parameters used in the training of the NNs presented in this work.}
    \begin{tabularx}{\columnwidth}{l|l}
    \hline
    \textbf{Number of hidden layers}  & 2                            \\ \hline
    \textbf{Neurons per hidden layer} & 50                           \\ \hline
    \textbf{Hidden layer activation}  & ReLU                         \\ \hline
    \textbf{Loss function}            & MSRE (Eq.~\ref{eq:loss})     \\ \hline
    \textbf{Optimizer}                & Adam                         \\ \hline
    \textbf{Training dataset size}    & $10^{6}$                     \\ \hline
    \textbf{Batch size}               & $32$                         \\ \hline
    \textbf{Number of epochs}         & 200                          \\ \hline
    \textbf{Initial learning rate}    & $10^{-3}$                    \\ \hline
    \end{tabularx}
  \label{tab:hyper-parameters}
\end{table}

\subsection{Neural network architectures}

\begin{figure*}[htb]
  \centering
  \begin{overpic}[width =.90\textwidth]{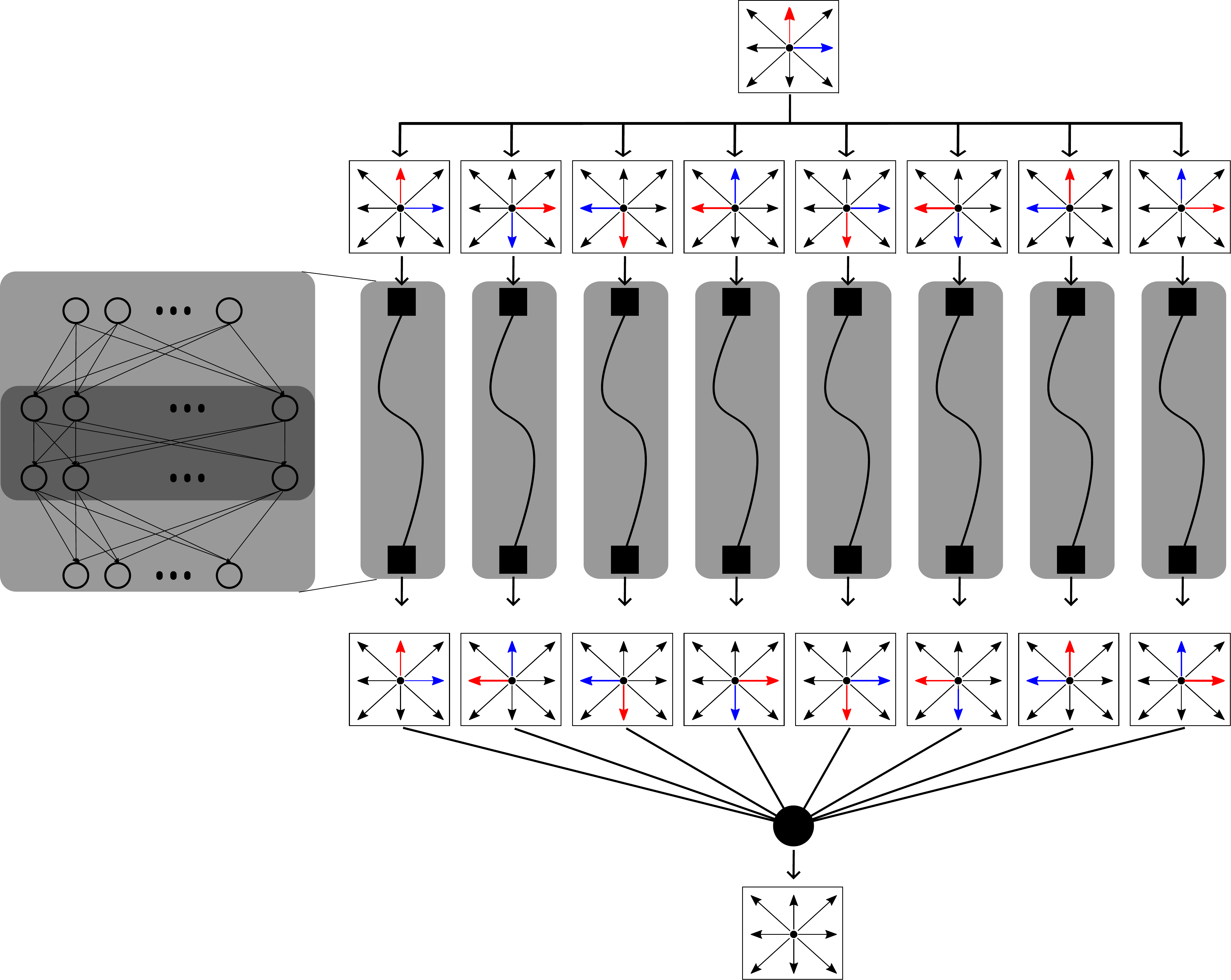}
    \put(70, 74){\color{black} \Large $f^{\rm pre }$ }
    \put(70,  3){\color{black} \Large $\tilde{f}^{\rm post}$ }
    \put(30, 70.2){\color{black}        $I     f^{\rm pre } $  }
    \put(40, 70.2){\color{black}        $r     f^{\rm pre } $  }
    \put(48, 70.2){\color{black}        $r^2   f^{\rm pre } $  }
    \put(56, 70.2){\color{black}        $r^3   f^{\rm pre } $  }
    \put(67, 70.2){\color{black}        $    s f^{\rm pre } $  }
    \put(75, 70.2){\color{black}        $r   s f^{\rm pre } $  }
    \put(83, 70.2){\color{black}        $r^2 s f^{\rm pre } $  }
    \put(93, 70.2){\color{black}        $r^3 s f^{\rm pre } $  }
    \put(3.5, 58.5){\color{black} \large $f_0^{\rm pre } $  }
    \put(8.5, 58.5){\color{black} \large $f_1^{\rm pre } $  }
    \put(17 , 58.5){\color{black} \large $f_8^{\rm pre } $  }
    \put(3.5, 28.5){\color{black} \large $\tilde{f}_0^{\rm post } $  }
    \put(8.5, 28.5){\color{black} \large $\tilde{f}_1^{\rm post } $  }
    \put(17 , 28.5){\color{black} \large $\tilde{f}_8^{\rm post } $  }   
    \put(98, 54){\color{black}        input   }
    \put(98, 33){\color{black}        output  }
    \put(30  , 28.8){\color{black}        $I            f^{\rm post } $  }
    \put(37  , 28.8){\color{black}        $r^{\text{-}1}       f^{\rm post } $  }
    \put(46.5, 28.8){\color{black}        $r^{\text{-}2}       f^{\rm post } $  }
    \put(56  , 28.8){\color{black}        $r^{\text{-}3}       f^{\rm post } $  }
    \put(65  , 28.8){\color{black}        $s^{\text{-}1}       f^{\rm post } $  }
    \put(72.5, 28.8){\color{black}        $(r   s)^{\text{-}1} f^{\rm post } $  }
    \put(82.5, 28.8){\color{black}        $(r^2 s)^{\text{-}1} f^{\rm post } $  }
    \put(93.5, 28.8){\color{black}        $(r^3 s)^{\text{-}1} f^{\rm post } $  }       
    \put(102, 62){\color{black}  \Large   $\sigma$         }
    \put(102, 43){\color{black}  \Large   $\Omega^{\rm NN}$}
    \put(102, 24){\color{black}  \Large   $\sigma^{-1}$    }    
    \put(25.5, 41  ){\rotatebox{90}{\color{black}        hidden  }}
    \put(27.0, 41.5){\rotatebox{90}{\color{black}        layers  }}
    \put(67, 11.8){\color{black} \Large  $\frac{1}{|D_8|} \sum $  }

  \end{overpic} 
  \caption{Sketch of a Neural Network architecture implementing the group averaging method. 
           The core network (gray box on the left hand side) is evaluated 8 times on 
           rotated/shifted versions of the input. The inverse transformation is applied to
           the 8 outputs which are then averaged in order to produce the final prediction.
          }\label{fig:network}
\end{figure*}
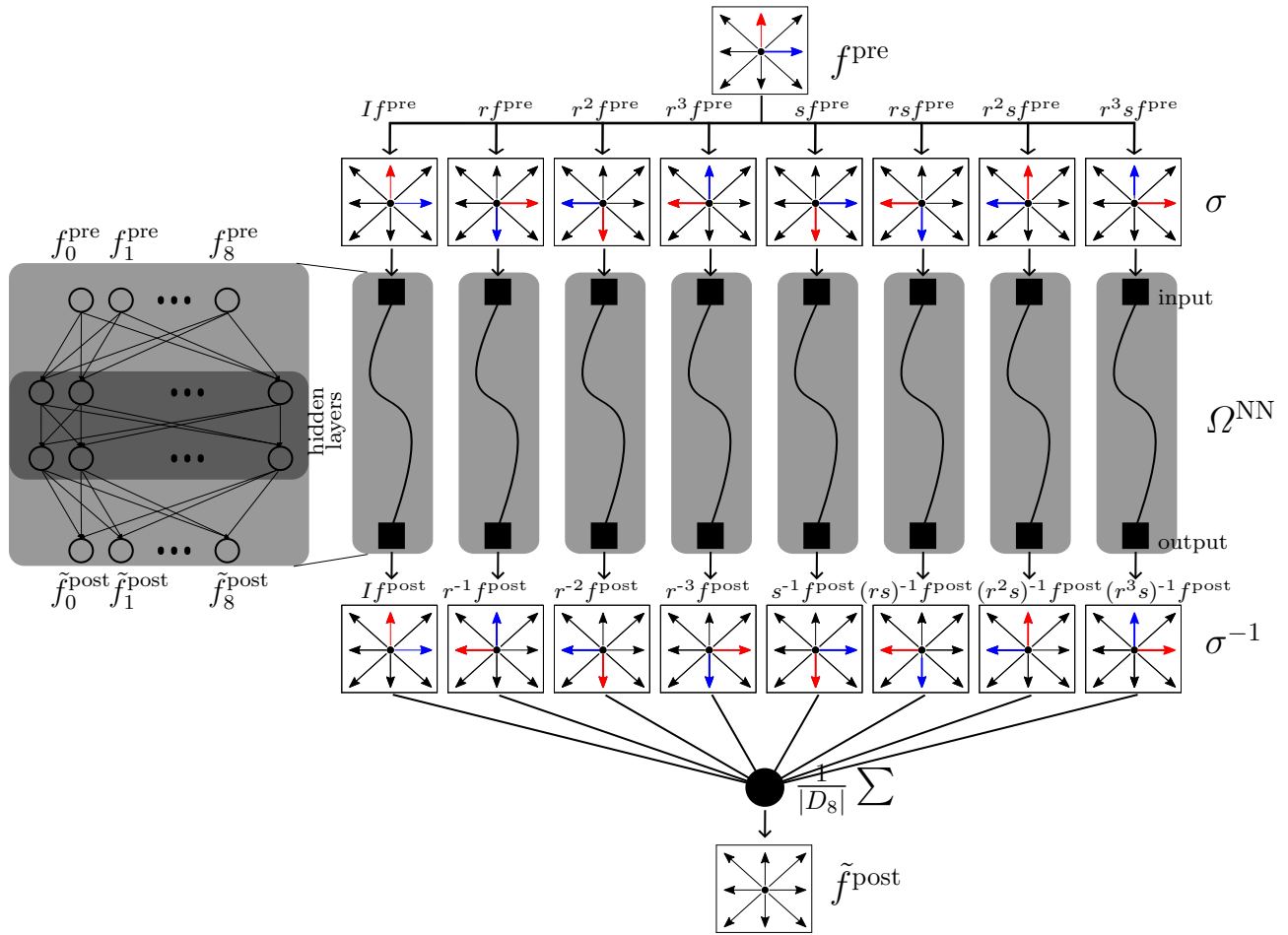

We consider variations of a fully connected feed-forward Neural Network, 
henceforth referred to as \textit{NN Naive}, which is composed of two hidden layers 
of 50 neurons each. We use ReLU as activation functions
and no biases in the linear layers.

The Naive NN, as it concatenates bias-less linear layers and ReLU activations, 
all degree-1 homogenous functions, is itself degree-1 homogeneous. 
Therefore it is hardwired to respect the scale equivariance P1.  
Yet, no other properties such as conservation of mass, momentum and 
$D_8$ equivariance are imposed, thus the denomination naive.

To amend this lack, in the reminder of this section we consider three 
further architectures:
\begin{itemize}
    \item \textit{NN Sym}, satisfying  properties P1, P2, P4;
    \item \textit{NN Cons}, satisfying  properties P1, P3;
    \item \textit{NN Sym+Cons}, satisfying properties P1, P2, P3.
\end{itemize}

Before detailing the structure of these networks, we present a more general 
approach to satisfy P1, which we will use in all next three architectures. 
It hinges on considering pre- and post-collision populations normalized by 
the corresponding macroscopic density (invariant, P3). 
In formulas, we effectively consider and train a NN, $\hat\Omega^{NN}$, 
operating as
\begin{equation}
  \tilde{\phi}_i^{\rm{post}} 
  = 
  \hat{\Omega}^{\rm{NN}}(\phi_i^{\rm{pre}}) ,
\end{equation}
where the normalized pre-collision populations are defined as
\begin{equation}\label{eq:pre-collision-normaliz}
    \phi_i^{\rm pre} 
    = 
    f_i^{\rm{pre}} / \rho 
    = 
    f_i^{\rm{pre}} / \sum_{i=0}^8 f_i^{\rm{pre}}. 
\end{equation}
The normalized post-collision populations are defined analogously.

Our final collision approximator, $\Omega^{\rm{NN}}$, 
prepends and appends rescaling operations as 
\begin{equation}\label{eq:nn-scale-invariant}
    \tilde{f}_i^{\rm{post}} 
    = 
    \Omega^{NN}(f^{\rm{pre}}_i) 
    = 
    \rho\hat\Omega^{NN}(\phi_i^{\rm{pre}}).
\end{equation}
On this basis, we can enforce positivity, P4,
by considering a softmax activation function 
at the final layer of the network (i.e., in place of a ReLU activation). 
Let $y_0,\ldots,y_8$ be the 9 inputs of the final activation, 
then the softmax outputs read
\begin{equation}\label{eq:softmax}
  \tilde{\phi}_i^{\rm{post}} 
  =  
  \frac{e^{y_i}}{Z}=\frac{e^{y_i}}{\sum_{i=0}^8 e^{y_i}} \ .
\end{equation}
Note that this returns normalized populations by construction 
(cf. Eq.~\ref{eq:pre-collision-normaliz}).

\subsection{$D_8$ equivariance: \textit{NN Sym}}\label{sec:symmetries}

We establish a collision NN, $\bar\Omega^{\rm{NN}}$, in which we enforce the rotation 
and symmetry equivariance (cf. Eq.~\ref{eq:P2}). 
We achieve this by applying a $D_8$ group averaging operation on 
a generic collision $\Omega^{\rm{NN}}$. 
In formulas, $\bar\Omega^{\rm{NN}}$ operates as follows
\begin{equation}\label{eq:group-averaging}
    \tilde{f}_i^{\rm{post}} 
    = 
    \bar\Omega^{\rm{NN}}(f_i^{\rm{pre}}) 
    =
    \frac{1}{|D_8|} 
    \sum_{\sigma \in D_8}\sigma^{-1}\Omega^{\rm{NN}}(\sigma f_i^{\rm{pre}}).
\end{equation}
A proof that Eq.~\ref{eq:group-averaging} satisfies P2 (Eq.~\ref{eq:P2}) 
is provided in Appendix~\ref{app:group-averaging}. 
Note that this approach is general: given any symmetry group the average 
in Eq.~\ref{eq:group-averaging} generates an operator that is equivariant 
with respect to such a group action. 
Note that here we perform a convex combination of populations, 
hence ensuring positivity of populations, 
with combined weight of unity, 
which ensures preservation of density
(assuming the original operator $\Omega^{\rm{NN}}$ had these properties).

In Fig.~\ref{fig:network}, we report our implementation of Eq.~\ref{eq:group-averaging}. 
Both at training time and for predictions the core network $\Omega^{\rm{NN}}$ 
is evaluated 8 times on rotated/shifted versions of the input ($\sigma f_i^{\rm{pre}}$). 
The outputs are then averaged 
after an application of the inverse rotation/shift ($\sigma^{-1}$). 

\subsection{Conservation of mass and momentum: \textit{NN Cons}}\label{sec:conservation}

A possible approach to ensure that Eq.~\ref{eq:conservation-laws} is satisfied,
is algebraically correcting the lattice populations 
which the NN outputs (see also Ref~\cite{miller-jocp-2022} for an example 
where hard-constraints on conservation laws are imposed on the full Boltzmann equation). 
The method is based on the observation that all the 
conserved quantities are linear combinations of the lattice populations.
Let 
\begin{equation}
    \bm{f} = [f_0,\dots,f_8]^T
\end{equation}
be the vector of the lattice populations, and $\bm{C}$ be an invertible matrix 
(representing change of bases):
\begin{equation}
    \bm{C} = [\bm{c}_0,\dots,\bm{c}_8]^T
\end{equation}
with
\begin{equation}
  \begin{aligned}
    \bm{c}_0 \cdot \bm{f} &= \rho \\ 
    \bm{c}_1 \cdot \bm{f} &= u_x  \\ 
    \bm{c}_2 \cdot \bm{f} &= u_y . \\ 
  \end{aligned}
\end{equation}
Consequently, the remaining column vectors $\bm{c}_3,\dots,\bm{c}_8$ are linearly independent and 
complementing $\bm{c}_0,\bm{c}_1,\bm{c}_2$ to a base of $\mR^9$. 

The matrix $\bm{C}$ represents an invertible map $\mR^9\rightarrow \mR^9$ 
which can be used to express a change of basis:
\begin{equation}
    \bm{b} = \bm{C} \bm{f} 
    \Longleftrightarrow 
    \bm{f} = \bm{C}^{-1} \bm{b}.
\end{equation}
Thus, the first three entries of $\bm{b}$ are the density and the momentum components.

Let $\bm{I}_1$ and $\bm{I}_2$  be two diagonal matrices 
adding up to identity matrix (i.e., $\bm{I}_1 + \bm{I}_2 = \bm{I}$), and satisfying
\begin{equation}
  \begin{aligned}
    \bm{I}_1 &= \textrm{diag}(1, 1, 1, 0, \dots, 0)  \\
    \bm{I}_2 &= \textrm{diag}(0, 0, 0, 1, \dots, 1)  \ .
  \end{aligned}
\end{equation}
We define the algebraic corrections as 
\begin{equation}\label{eq:alg-reconstruction}
  \begin{aligned}
    \tilde{\bm{f}}^{\rm{post}} 
    = 
    \Omega_c(\bm{f}^{\rm{pre}}) 
    =
    \bm{A} \bm{f}^{\rm{pre}} + \bm{B} \Omega^{\rm{NN}}(\bm{f}^{\rm{pre}}), \\
    \ \ \text{with} \quad
    \bm{A}=\bm{C}^{-1}\bm{I}_1\bm{C} \quad \text{and} \quad
    \bm{B}=\bm{C}^{-1}\bm{I}_2\bm{C}.
  \end{aligned}
\end{equation}
The choice of $\bm{A}$ and $\bm{B}$ is not unique. 
In what follows we will report results where the algebraic reconstruction
is applied to the populations of index $2, 5$ and $8$, using:
\def\hsp{\hspace{2.6mm}}
\begin{align*}
  \bm{A}
  & = 
  \left[
  \begin{array}{rrrrrrrrr}
   \hsp 0 & \hsp 0 & \hsp 0 & \hsp 0 & \hsp 0 & \hsp 0 & \hsp 0 & \hsp 0 & \hsp 0\\
   0 & 0 & 0 & 0 & 0 & 0 & 0 & 0 & 0 \\
   1 & 0 & 1 & 2 & 1 & 0 & 2 & 2 & 0 \\
   0 & 0 & 0 & 0 & 0 & 0 & 0 & 0 & 0\\
   0 & 0 & 0 & 0 & 0 & 0 & 0 & 0 & 0\\
   -\frac{1}{2} & \frac{1}{2} & 0 & -\frac{3}{2} & -1 & 1 & -1 & -2 & 0 \\
   0 & 0 & 0 & 0 & 0 & 0 & 0 & 0 & 0\\
   0 & 0 & 0 & 0 & 0 & 0 & 0 & 0 & 0 \\
   \frac{1}{2} & \frac{1}{2} & 0 & \frac{1}{2} & 1 & 0 & 0 & 1 & 1 \\
  \end{array}
  \right] 
  ,
  \\
  \bm{B} 
  & = 
  \left[
  \begin{array}{rrrrrrrrr}
   \hsp 1 & \hsp 0 & \hsp 0 & \hsp 0 & \hsp 0 & \hsp 0 & \hsp 0 & \hsp 0 & \hsp 0\\
   0 & 1 & 0 & 0 & 0 & 0 & 0 & 0 & 0 \\
  -1 & 0 & 0 &-2 &-1 & 0 &-2 &-2 & 0 \\ 
   0 & 0 & 0 & 1 & 0 & 0 & 0 & 0 & 0 \\
   0 & 0 & 0 & 0 & 1 & 0 & 0 & 0 & 0 \\
  \frac{1}{2} & -\frac{1}{2} & 0 & \frac{3}{2}  & 1 & 0 & 1 & 2 & 0 \\ 
   0 & 0 & 0 & 0 & 0 & 0 & 1 & 0 & 0 \\
   0 & 0 & 0 & 0 & 0 & 0 & 0 & 1 & 0 \\
  -\frac{1}{2} & -\frac{1}{2} & 0 & -\frac{1}{2} & - 1 & 0 & 0 & -1 & 0
  \end{array}
  \right] 
  .
\end{align*}

A second example is provided in Appendix~\ref{app:hard-const}.
Note that this approach allows to enforce the conservation of mass and momentum
at training time and yields no additional hyperparameters to be tuned.

An alternative approach, commonly adopted in the literature
~\cite{brunton-ams-2022,dener-arxiv-2020,byungsoo-cgf-2019},
consists of introducing a soft constraint in the loss 
function in order to penalize mass and momentum mismatches. 
In formulas, this reads:
\begin{equation}
  \mathcal{L} = \textrm{MSRE} + \alpha_1  | \tilde{\rho  } - \rho    | 
                              + \alpha_2 \| \bm{\tilde{u}} - \bm{u} \| \ ,
\end{equation}
where $\tilde{\rho}$ and $\bm{\tilde{u}}$ are the macroscopic quantities calculated over
the lattice populations output of the network $\tilde{f_i}^{\rm post}$,
while $\alpha_1$ and $\alpha_2$ weights the relative importance of each single constraint. 

Since we have observed that the imposition of hard constraints via algebraic reconstruction
systematically outperforms the soft-constraint based approach, the latter will not be 
covered in our analysis in the coming sections.
Nevertheless, a few numerical results are reported in Appendix~\ref{app:soft-const}
where we highlight the shortcomings of this approach.

\section{Numerical results}\label{sec:results}

\begin{figure}[h]
  \centering
  \begin{overpic}[width =.9\columnwidth]{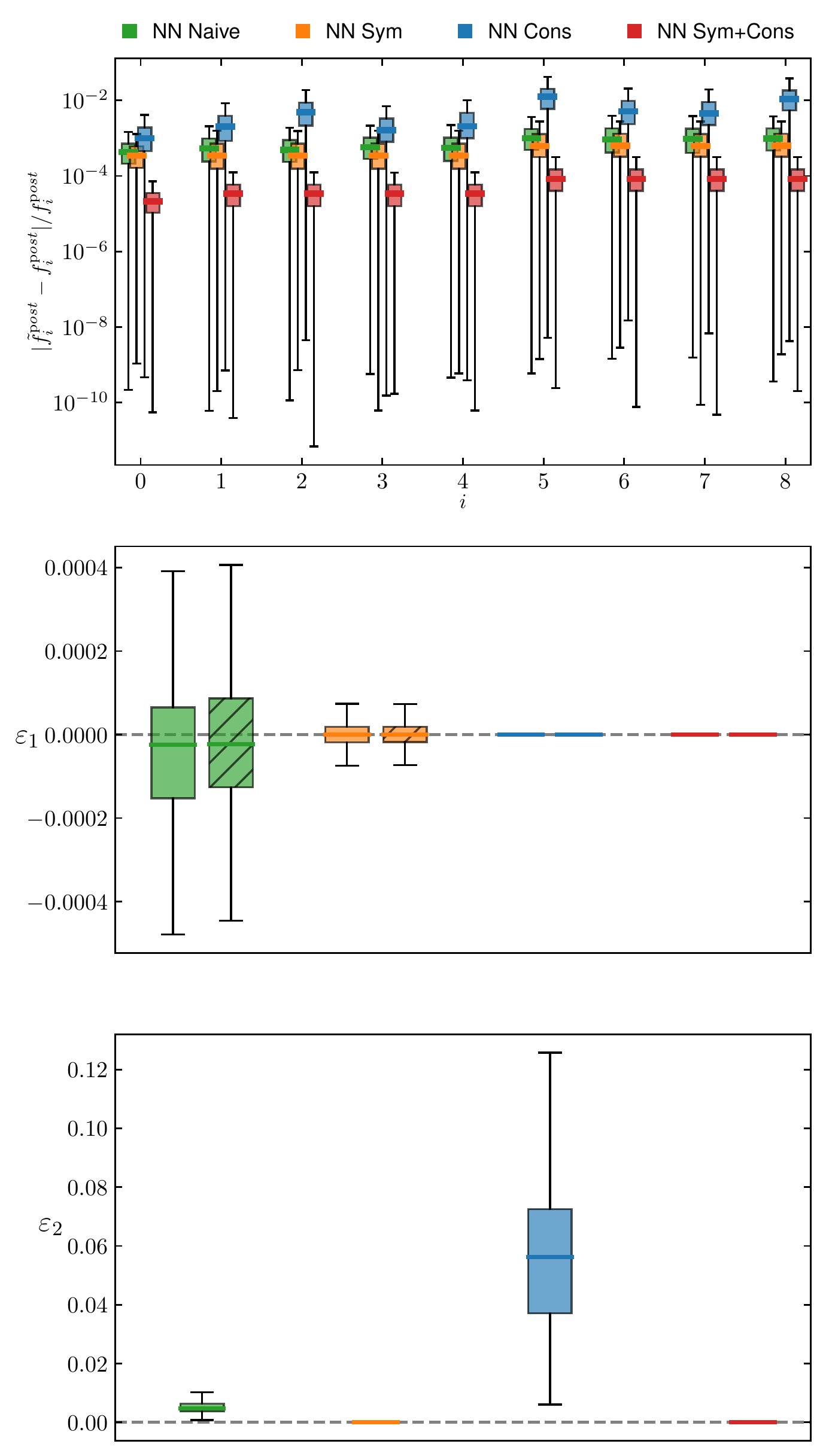}
    \put(  1, 98){\textbf{(a)}}
    \put(  1, 65){\textbf{(b)}}
    \put(  1, 33){\textbf{(c)}}
  \end{overpic}
  \caption{ Static evaluation of the accuracy achieved by the four different NN 
            architecture considered in this work.
            a) Comparison of the absolute relative error on the post-collision 
               populations of index $i$ (cf. Fig.~\ref{fig:d2q9}).
            b) Error committed in the conservation of momentum, with the uniformly filled 
               boxplots representing the error associated to $u_x$, and the boxplots with
               patterned filling the error associated to $u_y$ (see Eq.~\ref{eq:err1} 
               for the definition of the error metric). Note that the errors for 
               NN Cons and NN Sym+Cons are zero to machine precision.
            c) Error committed in violating rotation and mirroring equivariance 
               (see Eq.~\ref{eq:err2} for the definition of the error metric).
               Note that for NN Sym and NN Sym+Cons the error is zero down to machine precision.
          }\label{fig:static-evaluation}
\end{figure}

In this section, we present the results of LBM simulations where the collision term
is replaced by either of the four neural networks introduced in the previous section: 
NN Naive, NN Sym, NN Cons, NN Sym+Cons.  
For each NN architecture we trained 50 instances, adopting random weights initialization. 
We stop the training process at 200 epochs. See Table.~\ref{tab:hyper-parameters} for the
full list of training hyper-parameters.

We will first provide a static evaluation of the NN prediction error 
on the post-collision lattice 
populations. We also report on the physical properties of
the learned collision operator.
We will then turn our analysis to the comparison  of 
time dependent flows considering two standard benchmarks:  
a Taylor-Green vortex decay, and a lid-driven cavity flow.

\subsection{Static accuracy evaluation}

We start by comparing the accuracy of the various NN architectures described in 
the previous section taking into consideration the training error.
In Fig.~\ref{fig:static-evaluation}(a) we show the distribution of the
absolute relative error on the post-collision populations committed 
by the NN on the test dataset (generated following the procedure described 
in Appendix~\ref{app:training-set}).
The boxplots compare the accuracy of 50 different instances of each
NN architecture in the prediction of populations of index $i$.
By comparing the median values we observe that NN implementing 
symmetries slightly, although systematically, outperform the 
Naive NN.
On the other hand, hardwiring conservation laws does not lead 
to an improvement in the prediction of the lattice populations.
This is due to the specific choice of algebraically reconstructing
populations of index $2, 5$ and $8$ to restore the conservation 
of mass and momentum, and it can indeed be seen from the plot that the
largest errors area associated to these three elements.
A major improvement is achieved when combining conservation 
with rotation and symmetry equivariance (\textit{NN Sym+Cons}).
This case allows to improve accuracy in the prediction of 
the single lattice populations between 1 and 2 order of magnitudes
with respect to all the previous cases.

We now evaluate how well the different architecture comply to 
the physical properties of the collision operator.
In Fig.~\ref{fig:static-evaluation}(b) we evaluate the distribution of
the error committed in the momentum conservation by the various NN.
We define 
\begin{equation}\label{eq:err1}
  \varepsilon_1 = (u_j^{\rm pre} - u_j^{\rm post}) / c_s \ ,
\end{equation}
with $u_j^{\rm pre}$ the momentum calculated on the pre-collision distribution
functions, and $u_j^{\rm post}$ the momentum calculated from the
distribution functions predicted by the NN; in the plot 
the case $j = x$ is represented by the boxplots with uniform filling, and $j = y$
by the boxplots with patterned filling.

The $\varepsilon_1$ error distribution for the Naive NN 
is different when comparing the two spatial components, and also
asymmetric with respect to zero.
We observe that the NN implementing the symmetries of the lattice 
(\textit{NN Sym}) outperforms the Naive NN, in turn restoring 
the symmetry in the error distribution.
By construction, the error for the NN implementing conservation laws is 
systematically zero to machine precision.

\begin{figure}[htb]
  \centering
  \includegraphics[width =.99\columnwidth]{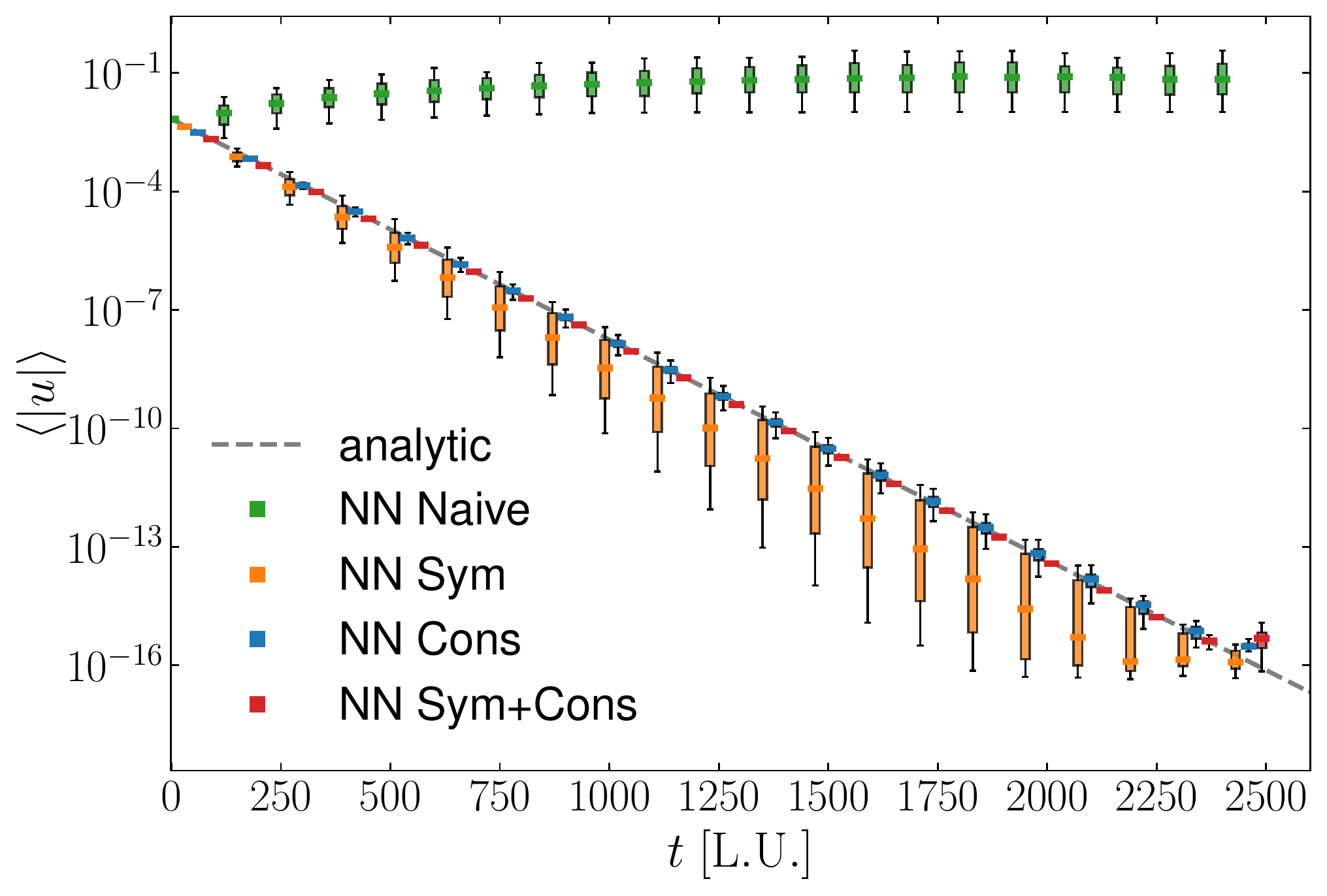}
  \caption{ Time evolution of the average absolute value of the velocity field 
            in a Taylor-Green vortex, comparing the analytic solution (gray dotted line)
            against simulations making use of NNs with different architectures.
            The boxplots
            show variability among 
            50 different instances 
            for each different NN architecture. 
            The NN with built-in symmetries and conservation properties (red)
            is the most accurate, followed by NN with only conservation properties (blue),
            followed by NN with only symmetries (orange).
            The naive NN (green) is the least accurate.
          }\label{fig:tg_decay_cmp}
\end{figure}

\begin{figure*}[htb]
  \centering
  \includegraphics[width =.95\textwidth]{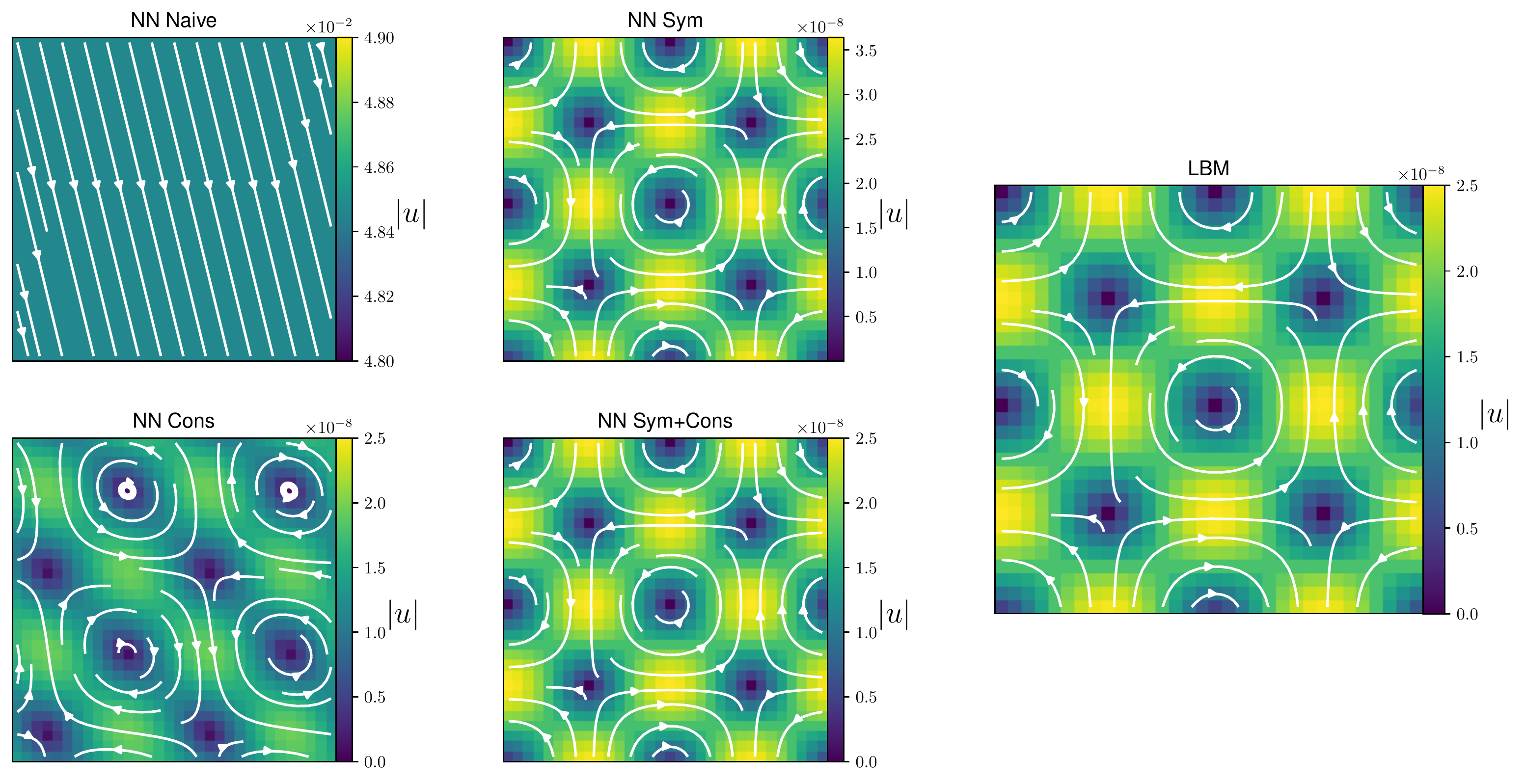}
  \caption{ Velocity profile from simulations of a Taylor-Green vortex decay, 
            after 1000 time steps. Colors map indicates the absolute value of the velocity vector,
            whereas white lines provide the velocity streamlines.
            We compare the ground truth from a LBM simulation against the results
            provided by different NN implementations.
          }\label{fig:tg_decay_snap}
\end{figure*}

Finally, in Fig.~\ref{fig:static-evaluation}(c) we evaluate the distribution 
of the following error metric
\begin{equation}\label{eq:err2}
  \varepsilon_2 = \frac{1}{|D_8|}\sum_{i=0}^8 \sum_{\sigma \in D_8}
   \Big|\sigma\Omega(f_i^{\rm{pre}}) 
  - 
  \Omega( \sigma f_i^{\rm{pre}} )\Big| \ ,
\end{equation}
which quantifies the violation of the $D_8$ equivariance. 
For $D_8$-equivariant collisions, i.e. satisfying P2 (Eq.~\ref{eq:P2}), 
the term within the absolute value is zero to machine precision. 
We evaluate $\varepsilon_2$ over the entire test dataset.
We observe that the network implementing conservation laws 
(\textit{NN Cons}) commits a larger error even when comparing with the Naive NN.
This is due to the fact that the algebraic reconstruction procedure used to 
implement the conservation laws leads to the error accumulating along some 
lattice directions.
The error metric is systematically zero for all the NN implementing the 
group-averaging technique.

In the coming sections we compare the performance of the different NN
in the simulation of time-dependent fluid flows.

\subsection{Benchmark I: Taylor-Green Vortex}

We consider the time dynamics of a Taylor-Green vortex, 
a standard benchmark for the validation of fluid flow solvers since it 
provides an exact solution to the Navier-Stokes equations.

Starting from the following initial conditions in a 2D periodic domain:
\begin{equation}
  \begin{aligned}\label{eq:tg-initial-conditions}
    u_x(x,y) &=& u_0 \cos{\left(x\right)} \sin{\left(y\right)} &, \\
    u_y(x,y) &=&-u_0 \cos{\left(y\right)} \sin{\left(x\right)} &, \quad x,y \in [0, 2 \pi]
  \end{aligned}
\end{equation}
with $u_0$ the initial value for $|\bm{u}|$,
it is simple to show that the flow decays 
exponentially and proportionally to
\begin{equation}
  F(t) = \exp{\left( -2 \nu t \right)}  ,
\end{equation}
where $\nu$ is the kinematic viscosity of the fluid (Eq.~\ref{eq:viscosity}).
This benchmark allows us to evaluate the time dynamic of a flow, 
covering different orders of magnitude in the values of the macroscopic parameters,
and also to evaluate the preservation of symmetries by observing 
the structure of the vortexes. 

We consider a $32 \times 32$ grid, with $u_0 = 10^{-2}$, $\tau = 1$.
In Fig.~\ref{fig:tg_decay_cmp} we compare the time decay of the average absolute
value of the velocity field from simulations making use of different NNs,
comparing against the analytic solution. Once again, for each type 
of NN we have evaluated the results from 50 different networks trained 
starting from a random choice of the initial weights. 
The plot highlights the variability in the 
results from the different NNs by means of boxplots.
From the plot we can see that the Naive NN is able of correctly follow the 
flow decay for only 20-40 iterations, after which not only the flow stops decaying
but we also observe an increase in the kinetic energy.
By employing a NN satisfying the symmetries of the lattice it is possible to
restore the decaying trend of the flow, although we observe a deviation
from the correct decaying rate. This can be attributed to the network not being
able of preserving momentum.
On the other hand, NNs enforcing the conservation laws are able to 
provide a more accurate dynamic, with only small variability around 
the analytic solution, which can be further reduced by 
combining conservation and preservation of symmetries.

\begin{figure}[htb]
  \centering
  \begin{overpic}[width =.95\columnwidth]{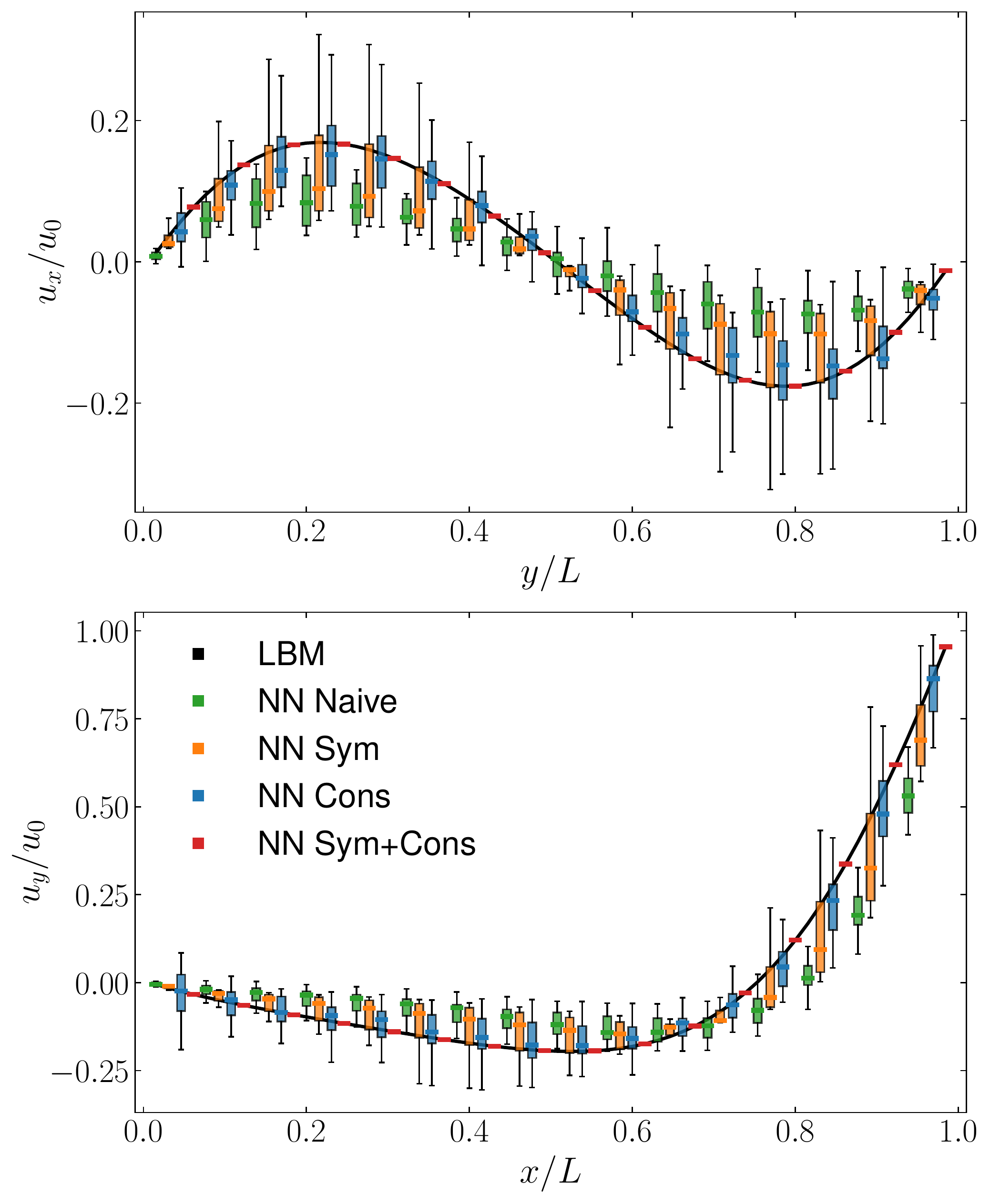}
    \put(  1, 98){\textbf{(a)}}
    \put(  1, 50){\textbf{(b)}}
  \end{overpic}  
  \caption{ Steady state profiles for (a) $u_x$ along the vertical centerline, 
            and (b) $u_y$ along the horizontal centerline of the domain of a 
            lid-driven cavity flow at $\rm{Re} = 10$.
            Simulations are performed on a square grid of side $L = 64$.
            We compare the results of a LBM simulation (black line), against 
            results obtained employing the four NN architectures considered in 
            this work. The boxplots report the variability in the results from
            50 instances of each NN architecture. 
          }\label{fig:ldcf_cmp_Re10}
\end{figure}

\begin{figure}[htb]
  \centering
  \begin{overpic}[width =.95\columnwidth]{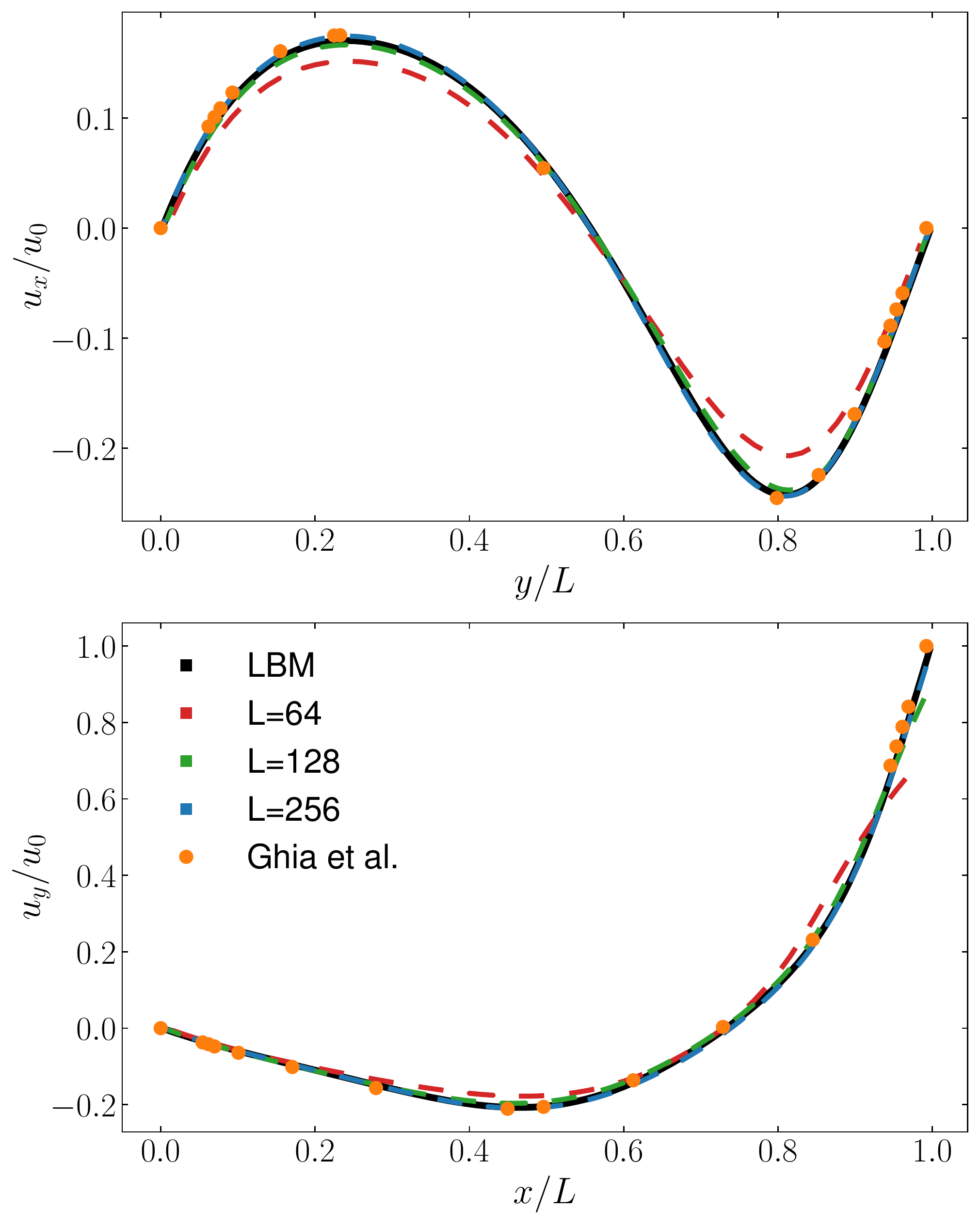}
    \put(  1, 98){\textbf{(a)}}
    \put(  1, 50){\textbf{(b)}}
  \end{overpic}    
  \caption{ Steady state profiles for (a) $u_x$ along the vertical centerline, 
            and (b) $u_y$ along the horizontal centerline of the domain of a lid-driven
            cavity flow at $\rm{Re} = 100$.
            Dotted lines represent results obtained using a \textit{NN Sym+Cons}
            architecture for increasing number of nodes in the grid side $L$.
            We compare the results against a LBM simulation (black line, $L = 256$),
            and reference data from Ref.~\cite{ghia-jcp-1982} (orange dots, $L = 129$).
          }\label{fig:ldcf_cmp_Re100}
\end{figure}

\begin{figure*}[htb]
  \centering
  \includegraphics[width =.99\textwidth]{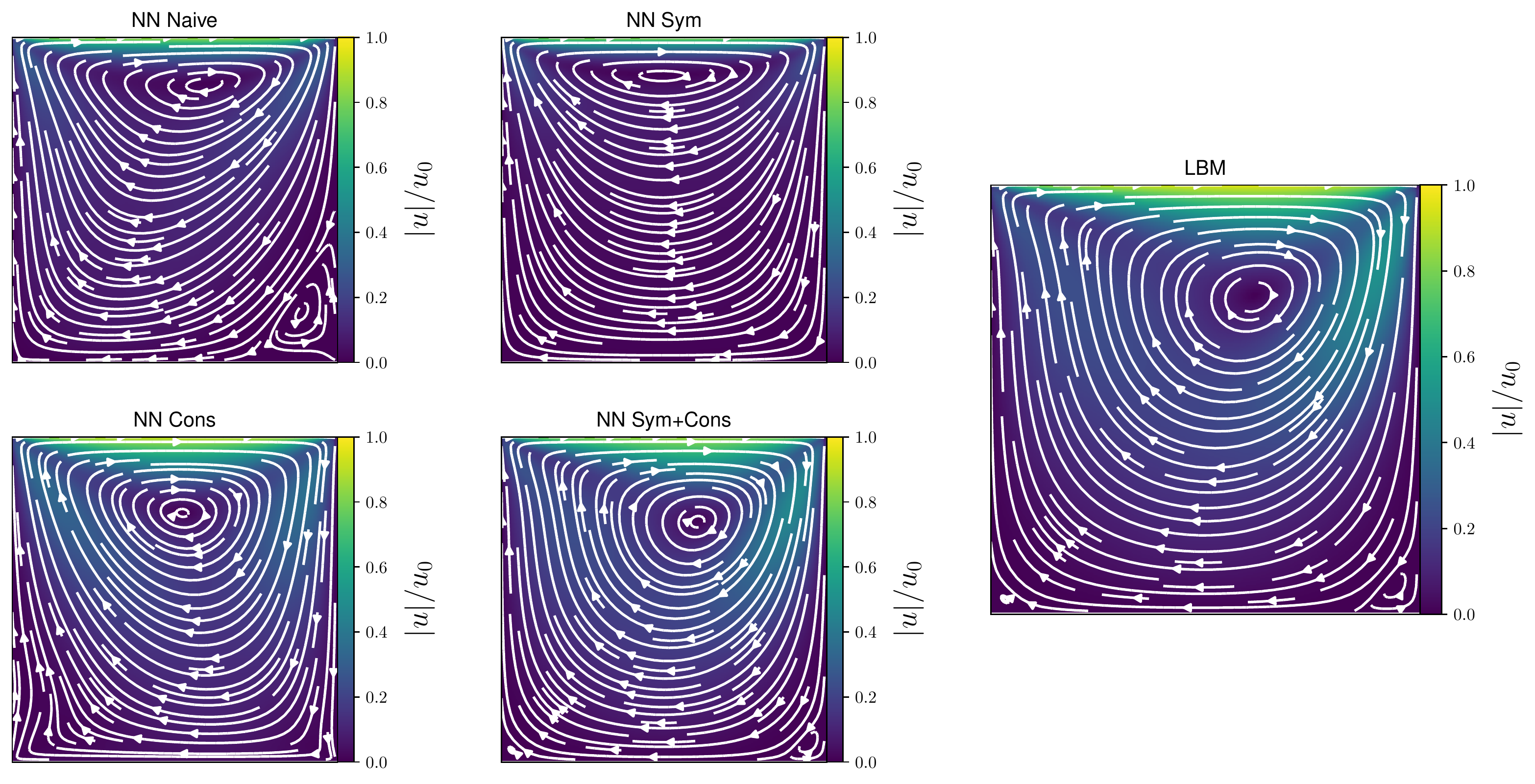}
  \caption{ Steady state profile of the velocity field for a lid-driven cavity flow
            at $\rm{Re} = 100$, comparing the results of a LBM simulation against
            the results provided by different NN implementations. Colors map the 
            absolute value of the fluid velocity normalized over the lid velocity,
            whereas white lines provide the velocity streamlines. Simulations
            are performed on a square grid of side $L = 128$.
          }\label{fig:ldcf_snap}
\end{figure*}

The importance of embedding conservation laws and symmetries together in the NN 
is even more evident in Fig.~\ref{fig:tg_decay_cmp},
where evolution statistics is shown for four types of NN designs. 
Embedding symmetries or conservation properties shows an immediate and dramatic improvement over the naive NN 
in the ability of the NN to capture the decay rate of the average velocity field.
Enforcing conservation properties is appreciably more important (for the purpose of this test) than enforcing symmetries.
Yet, enforcing both symmetries and conservation properties produces the most accurate 
results capturing the decay of average velocity with minimal variability all the way 
to machine precision, which is a remarkable result, especially compared to the performance of a naive NN. 
Moreover, we should stress that a NN with a lower training error will 
not necessarily guarantee for better results when employed in simulations;
for example, NN Cons, which in Fig.~\ref{fig:static-evaluation}(a) 
presents the larger training error, is among the best performing one
when looking at Fig.~\ref{fig:tg_decay_cmp}.

On a more qualitative basis, in Fig.~\ref{fig:tg_decay_snap} we provide snapshots of the velocity field 
at a later stage of the dynamics (after $t = 1000$ iterations),
comparing the ground truth given by a plain LBM simulation against an 
example of the profile provided by each of the different NN implementations.
The figure shows that, besides failing to reproduce the decay of the flow, the 
Naive NN is also not able to preserve the structure of the vortexes. 
The NN with symmetries, on the other hand, nicely preserves the geometric structure,
although the amplitude of the velocity is slightly off with respect to the 
reference LBM profile.
The NN enforcing conservation laws correctly captures on average the decaying rate
(c.f. Fig.~\ref{fig:tg_decay_cmp}), however, Fig.~\ref{fig:tg_decay_snap} 
clearly shows that the structure of the vortexes is not symmetric anymore.
This can be attributed to the fact that the algebraic reconstruction is performed
on 3 lattice populations, leading to a less balanced distribution of the error 
(cf. Fig.~\ref{fig:static-evaluation}(c)).
Only by combining conservation and symmetries in the NN it is possible to reproduce
correctly the velocity profile.

\subsection{Benchmark II: Lid driven cavity flow}

As a second example, we consider the lid-driven cavity flow, a wall-bounded  benchmark 
in a very simple geometry, still leading to a non-trivial dynamic. Indeed there is no
analytic solution for this flow, and for this reason we will compare this time only
against reference LBM simulations.

The setup consists of a top-lid moving at a constant velocity ($u_0$), with no-slip
boundary conditions at bottom and side walls. 
We consider a $L \times L$ grid, the relaxation time set to $\tau = 1$, 
and report the results for simulations at two different Reynolds numbers, 
respectively $\rm{Re} = 10$ and $\rm{Re} = 100$, with
\begin{equation}\label{eq:reynolds}
  \rm{Re} = \frac{u_0 L}{\nu} \ .
\end{equation}

In simulations the NN does not handle the evolution of the boundary nodes.
Instead, we employ standard LBM approaches for implementing the boundary conditions. 
In particular, the bounce back rule is used to implement the no-slip condition.
Here the lattice populations that during the streaming step interact with a solid wall
get reflected to their original location with their velocity reversed: 
\begin{equation}
  f_{\bar{i}} (\bm{x}, t + 1) = f_i (\bm{x}, t) \ ,
\end{equation}
where $f_{\bar{i}}$ is the population of index $\bar{i}$ such that 
$\bm{\xi}_{\bar{i}} = - \bm{\xi}$.
For the top wall we use a Dirichlet boundary condition
\begin{equation}
  f_{\bar{i}} (\bm{x}, t + 1) 
  = 
  f_i (\bm{x}, t) + 2 w_i \rho_w \frac{\bm{\xi}_i \cdot \bm{u}_w}{c_s^2} \ ,
\end{equation}
where $\rho_w$ and $\bm{u}_{w} = (u_0, 0)$ are respectively the density and 
the velocity at the top wall.

In Fig.~\ref{fig:ldcf_cmp_Re10} we show the steady state velocity profiles 
along the vertical (a) and horizontal (b) centerlines of the lid-driven cavity
for $\rm{Re} = 10$, comparing the results from a plain LBM simulation 
against results obtained employing NNs with different architectures.
All simulations are performed on a square grid of side $L = 64$.
Once again we show data collected simulating 50 different instances of each 
NN architecture, with the boxplots reporting the variability in the obtained results.
We observe that in this case the results of the Naive NN are much closer to the
reference data with respect to the previous benchmark. This can be attributed to
the boundary conditions constraining the flow. 
Both \textit{NN Sym} and \textit{NN Cons} provide an improvement over the Naive NN,
however it is interesting to point out that the results provided by the latter 
present a much higher variability than the one observed in the simulation of the
Taylor-Green vortex.
Indeed, the plot clearly shows that only the case \textit{NN Sym+Cons} is able to 
correctly reproduce the results of the LBM simulation.
We select this NN architecture to perform simulation at a higher Reynolds number.
In Fig.~\ref{fig:ldcf_cmp_Re100} we show the results obtained at $\rm{Re} = 100$,
varying the grid size, and comparing with both a LBM simulation as well as with
reference data from Ghia et al.~\cite{ghia-jcp-1982}.
The results from the simulation using the finer grid resolution ($L=256$) are found 
to be in excellent agreement with the reference data. On the other hand, we see that
for coarser grid sizes the NN struggles to correctly reproduce the velocity in the
proximity of the moving plate (see Fig.~\ref{fig:ldcf_cmp_Re100}(b)).
We shall discuss the origin of this mismatch in the coming subsection.

In Fig.~\ref{fig:ldcf_snap} we show a more qualitative comparison for the
case $\rm{Re} = 100$, presenting snapshots of the velocity field at the steady state,
and comparing the results from a LBM simulation with results produced by the 
different NN architectures. 
It is interesting to observe that each different NN make a different prediction for 
the location of the main vortex, and only few reproduce the secondary vortex located 
at the bottom right corner. As expected from the analysis above, \textit{NN Sym+Cons}
provides results in excellent agreement with the LBM simulation.

\subsection{Extrapolation }

\begin{figure}[htb]
  \centering
  \includegraphics[width =.95\columnwidth]{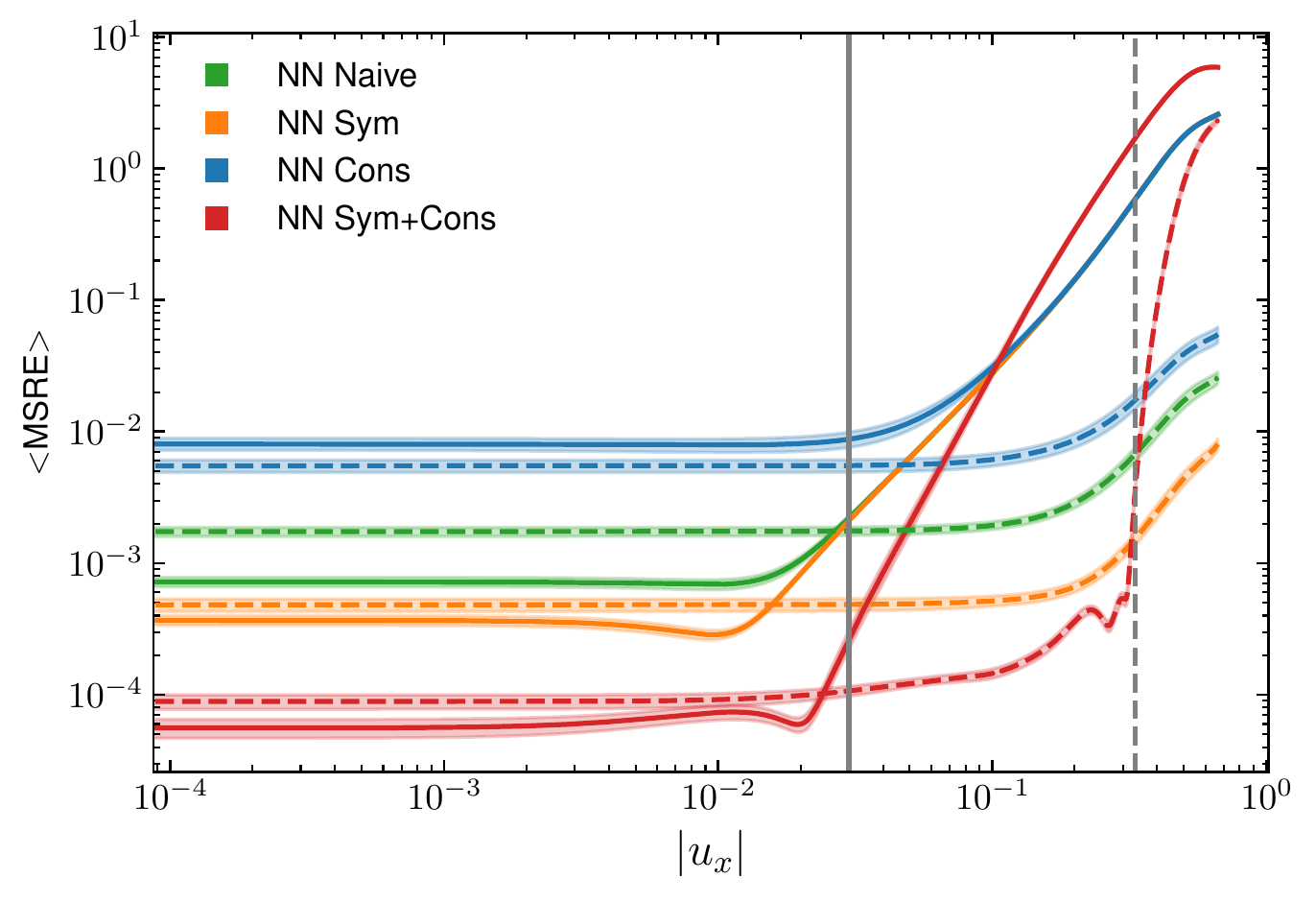}
  \caption{ Comparison for the accuracy of the different NNs architecture 
            within and outside the training dataset.
            The plot shows the average MSRE, computed on 50 instances of each NN 
            architecture, in the prediction of the equilibrium distribution 
            $f_i^{\rm eq}(\rho = 1, u_x, u_y=0)$ at varying values of $u_x$. 
            The continuous lines refer to NNs trained on a dataset where 
            the macroscopic velocity takes values in the interval $(-0.03, 0.03)$,
            while the interval $(-1/3, 1/3)$ has been used to train the NNs
            corresponding to the dotted lines. The gray continuous (dotted) 
            vertical lines identify the boundary of the two training datasets.
          }\label{fig:extrapolation-error}
\end{figure}

In Fig.~\ref{fig:ldcf_cmp_Re100}(b) we have observed significant deviations in the numerical
results produced by the \textit{NN Sym+Cons} architecture in the proximity of the 
moving plate, in particular for coarse grids. Since in simulations we are 
keeping fixed the kinematic viscosity and the Reynolds number, it follows from 
Eq.~\ref{eq:reynolds} that by increasing the grid resolution we also decrease the 
numerical value of the lid velocity $u_0$. 
For $L=64$ the numerical value used at the top lid $u_0 \approx 0.26$ falls well outside
the range of values shown to the network at training time.
It is therefore interesting to investigate the extrapolation capabilities of the different NNs.
In Fig.~\ref{fig:extrapolation-error} we show the average MSRE on 50 instances of each NN 
architecture, calculated in the prediction of the equilibrium distribution 
$f_i^{\rm eq}(\rho = 1, u_x, u_y=0)$ at varying values of $u_x$. 
The continuous lines show the performance of the NNs trained 
on a dataset where the macroscopic velocity takes values in the 
interval $(-0.03, 0.03)$; likewise, the dotted lines 
show the results for NNs trained on values of the macroscopic velocity
in the interval $(-1/3, 1/3)$. Corresponding gray continuous (dotted) 
vertical lines are reported to identify the boundary of the two training datasets.
Here we can see that when working in the range of values shown to the NN
during the training, the \textit{NN Sym+Cons} outperforms all the other
network architectures. On the other hand, this NN commits the largest
extrapolation error, i.e. it commits a larger error in predicting 
the equilibrium distribution outside of the values of the training set.
While the reason for this behavior is currently unclear to us and will be object
of further analysis in future work, these results explain the discrepancies 
observed in Fig.~\ref{fig:ldcf_cmp_Re100}, and in turn point to the need
of extra care in the preparation of the training dataset.

\section{Conclusion}\label{sec:conclusions}

In this work we have presented a machine learning approach for learning 
a collision operator for the Lattice Boltzmann Method from data.
As a proof of concept, we have developed a neural network capable of 
approximating to good accuracy the BGK collision operator.
We have discussed in details a few methods which allow enriching the 
structure of the neural network to enforce relevant physical
properties of the collision operator.
We have shown that only by embedding conservation laws and
lattice symmetries in the neural network it is possible to 
correctly reproduce the time dynamics of a fluid flow.

This work can be regarded as a first step towards the application of 
neural networks for extending the applicability of LBM in kinematic 
regimes not supported by the basic method.
To give an example, in future extensions of the present work,
we plan to evaluate the possibility of using our approach for learning
collision operators from molecular dynamics and Monte Carlo simulations
in regimes beyond hydrodynamic limit.

\begin{appendices}

\section{Training data generation algorithm}\label{app:training-set}

In this appendix section we summarize the steps followed in the generation of the 
training dataset. While the procedure described in Algorithm~\ref{alg:training-set} 
is general, we provide values which are specific for the D2Q9 model 
(for example the coefficients in Eq.~\ref{eq:fneq-correction}).

\begin{algorithm}[htb]
  \caption{Training data generation algorithm}\label{alg:training-set}
  \begin{algorithmic}[1]
    \State 
      Sample
      \begin{equation}
        \begin{aligned}
          \rho \log{\rho}  & \sim \mathcal{U}(\rho_{\rm min}, \rho_{\rm max}) \\
          \bm{u} & = u_0 (\cos(\theta), \sin(\theta)) \ ,
        \end{aligned}
      \end{equation}
      with 
      \begin{equation}
        \begin{aligned}
          u_0  \log{u_0}   & \sim \mathcal{U}(0, u_0^{\rm max}) \\
          \theta & \sim \mathcal{U}(0,2\pi) \ .
        \end{aligned}
      \end{equation}
    \State
      Compute 
      \begin{equation}
        f_i^{\rm{eq}} = f_i^{\rm{eq}}(\rho,\bm{u})
      \end{equation}
      using Eq.~\ref{eq:feq}.
    \State
      Sample
      \begin{equation}
        f^{\prime \rm{neq}}_i \sim \mathcal{N}(0, \sigma^2)
      \end{equation}
      from normal distribution $\mathcal{N}(0, \sigma^2)$ with mean 0 and standard deviation $\sigma$. 
    \State
      Map $f'^{\rm{neq}}_i$ to a mass- and momentum-less $f^{\rm{neq}}_i$ (Eq.~\ref{eq:fneq})
      \begin{equation}\label{eq:fneq-correction}
        f^{\rm{neq}}_i = f^{\prime \rm{neq}}_i - \frac{1}{9} \rho^{\prime} 
                                               - \frac{1}{6} \bm{\xi}_i \cdot \rho^{\prime} \bm{u}^{\prime}. 
      \end{equation}
      where
      \begin{equation}
        \begin{aligned}
          \rho^{\prime}                 &= \sum_{i=0}^{8} f^{\prime \rm{neq}}_i \\
          \rho^{\prime} \bm{u}^{\prime} &= \sum_{i=0}^{8} f^{\prime \rm{neq}}_i \bm{\xi}_i \ .
        \end{aligned}
      \end{equation}
    \State
      Compute 
      \begin{equation}
        f_i^{\rm{pre}} = f_i^{\rm{eq}} + f_i^{\rm{neq}}.
      \end{equation}
    \State
      Compute
      \begin{equation}
        f_i^{\rm{post}} = \Omega(f_i^{\rm{pre}})
      \end{equation}
      using the BGK collision operator in Eq.~\ref{eq:lbgk}.
  \end{algorithmic}
\end{algorithm}

\section{Conservation with soft-constraints }\label{app:soft-const}

In Section~\ref{sec:ml} we have discussed the possibility of employing soft-constraints
in order to impose conservation of mass and momentum.
In this appendix, we report the results obtained training a NN with the following architecture:
\textit{i)} same hyperpameters as in Table~\ref{tab:hyper-parameters}, 
\textit{ii)} softmax activation function at the final layer, 
combined with the rescaling operations in Eq.~\ref{eq:nn-scale-invariant},
\textit{iii)} an additional term in the loss function penalizing the violation of momentum conservation:
\begin{equation}\label{eq:momentum-soft-const}
  \mathcal{L} = \textrm{MSRE} + \alpha \frac{ \| \bm{\tilde{u}} - \bm{u} \| }{ \| 1 + \bm{u} \| }   ,
\end{equation}
where $\bm{\tilde{u}}$ is the velocity vector computed over the lattice populations output
of the NN, and $\alpha$ is a parameter which weights the importance of the soft constraint. 
Note that mass conservation is already ensured by the combination of the softmax activation
function with the rescaling of input and output.

\begin{figure}[htb]
  \centering
  \includegraphics[width =.99\columnwidth]{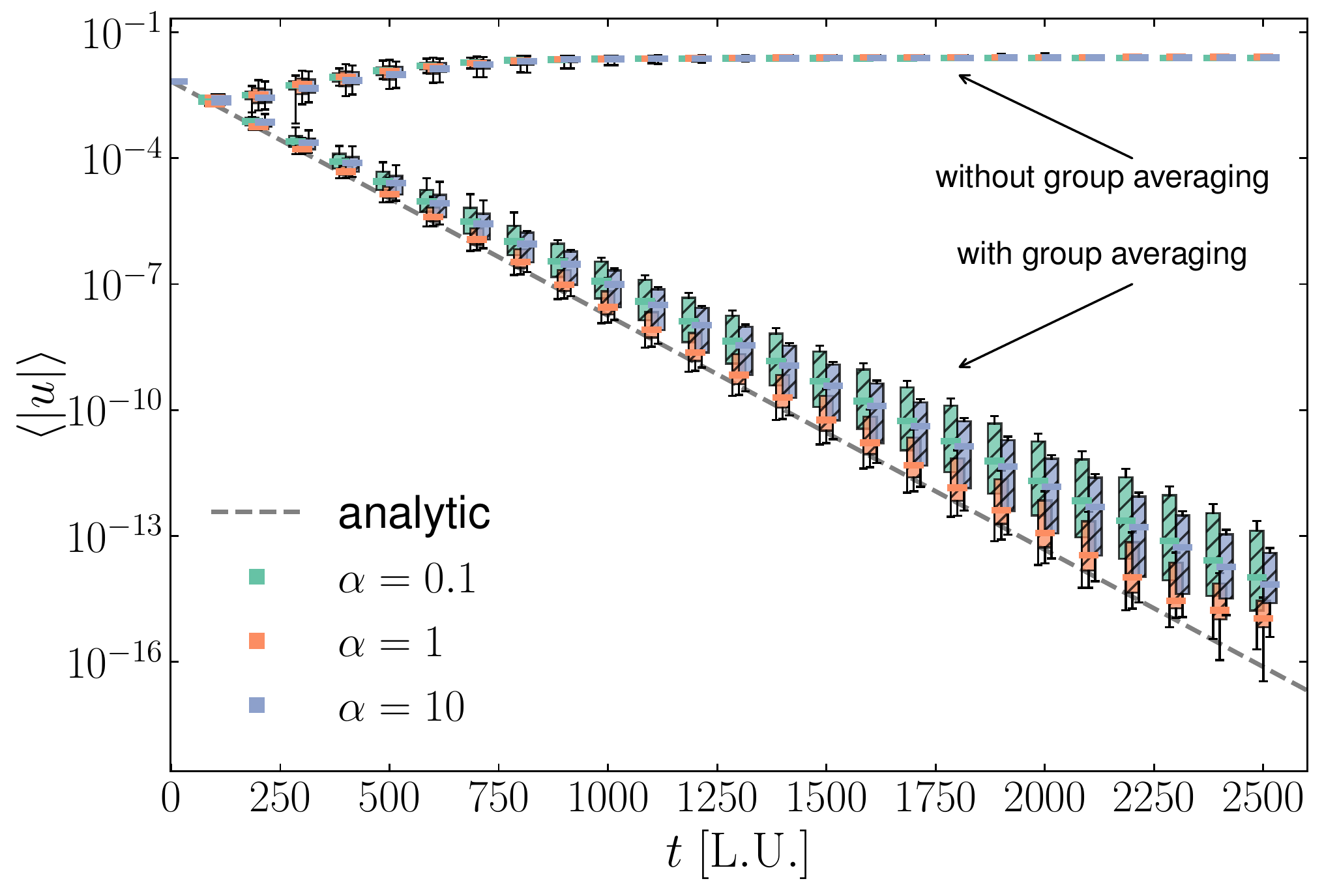}
  \caption{ Time evolution of the average absolute value of the velocity field 
            in a Taylor-Green vortex, comparing the analytic solution (gray dotted line)
            against simulations making use of NNs employing a soft-constraint on momentum 
            conservation, for a few selected values of the parameter $\alpha$ which
            weights the relative importance of the soft-constraint (see Eq.~\ref{eq:momentum-soft-const}).
            For each value of $\alpha$ we show results obtained with and without 
            the group-averaging method used for embedding symmetries in the NN architecture.
            The boxplots show variability among 20 different instances for each different NN architecture. 
          }\label{fig:tg_decay_softconst}
\end{figure}

We have scanned several values of the parameter $\alpha$, for which we report here
three representative cases: $\alpha = 0.1, 1, 10$. For each of these selected values
of $\alpha$ we have trained 20 NNs.
In Fig.~\ref{fig:tg_decay_softconst} we present the results obtained on the Taylor Green
vortex benchmark described in the main text.
For the case where no symmetries are enforced in the NN, the results are inline with
those reported for the Naive NN in Fig.~\ref{fig:tg_decay_cmp}, i.e. we do not 
correctly reproduce the decaying behavior if not for very few time steps.
By repeating the training embedding symmetries in the NN architecture (Fig.~\ref{fig:network}),
results improve significantly. From the plot we can observe that by tuning $\alpha$ it 
is possible to adjust the variability in the results produced by the different instances 
of the NN. Still, we not in general achieve the correct decaying rate.

These results show that NNs imposing conservation laws via hard constraints systematically
outperforms the soft-constraints based approach, with the added advantage of not requiring 
tuning of extra parameters (such as $\alpha$ in the example above).

\section{Symmetric algebraic reconstruction }\label{app:hard-const}

In the main text we have described one possible way to apply algebraic reconstruction 
to hardwire conservation laws in the NN. In particular we have considered an example
which involves adjusting 3 of the 9 populations outputted by the NN.
This approach, which may introduce a slight bias along those lattice directions, 
was useful to expose the relative importance of embedding different physical properties
in the NN architecture.

However, as mentioned in Section~\ref{sec:ml}, there are several possible approach 
for imposing conservation of mass and momentum in the NN.

A more ``symmetric'' approach, which we found to give excellent results even when 
not combined with the group-averaging method for embedding symmetries in the NN,
reads as follows:
\begin{equation} 
    \tilde{f_i}^{\rm{post}} 
    = 
    \Omega^{\rm NN}(f_i^{\rm{pre}}) + \kappa_1 + \kappa_2 \xi_{i, x} + \kappa_3 \xi_{i, y}
\end{equation}

where the parameters $\kappa_1, \kappa_2, \kappa_3$ are lattice dependent, and for the D2Q9 
are given by:
\begin{equation}
  \begin{aligned}
    \kappa_1 &=  - \frac{1}{9} \sum_{i=0}^8 \left(  \Omega^{\rm NN}(f_i^{\rm{pre}}) - f_i^{\rm{pre}} \right)            \\
    \kappa_2 &=  - \frac{1}{6} \sum_{i=0}^8 \left(  \Omega^{\rm NN}(f_i^{\rm{pre}}) - f_i^{\rm{pre}} \right) \xi_{i, x} \\
    \kappa_3 &=  - \frac{1}{6} \sum_{i=0}^8 \left(  \Omega^{\rm NN}(f_i^{\rm{pre}}) - f_i^{\rm{pre}} \right) \xi_{i, y} \\
  \end{aligned}
\end{equation}

\section{The group-averaged operator $\bar\Omega$ satisfies $D_8$ equivariance}\label{app:group-averaging}

We prove here that the the group averaged operator defined in Eq.~\ref{eq:group-averaging}, 
respects property P2 (Eq.~\ref{eq:P2}). 
The proof that we propose here is indeed general and holds for any symmetry group. 
For this reason we indicate here the symmetry group with the generic symbol $\Sg$.
\paragraph{Proof.} We shall show that 
\begin{equation*}
    \bar{\Omega}(\eta f) = \eta \bar{\Omega}(f),\quad \forall \eta \in \Sg.
\end{equation*}

\noindent By definition, it holds
\begin{align*}
    \bar{\Omega}(\eta f) 
    &= \frac{1}{|\Sg|}\sum_{\sigma \in \Sg} \sigma^{-1} \Omega^{\rm{NN}}(\sigma \eta f) \\
    &= \frac{1}{|\Sg|}\sum_{\sigma \in \Sg}\eta\eta^{-1} \sigma^{-1} \Omega^{\rm{NN}}(\sigma \eta f) \\
    &= \eta \frac{1}{|\Sg|}\sum_{\sigma \in \Sg}(\sigma\eta)^{-1}  \Omega^{\rm{NN}}(\sigma \eta f) \\
    &= \eta \frac{1}{|\Sg|}\sum_{\sigma \in \Sg\eta}\sigma^{-1}  \Omega^{\rm{NN}}(\sigma f).
\end{align*}
Yet, $\Sg\eta = \Sg$ or, equivalently, the presence of $\eta$ yields 
a permutation of the terms to add  (by uniqueness of the inverse within a group). 
Thus, we can write
\begin{align*}
    \bar{\Omega}(\eta f) 
    &= \eta \frac{1}{|\Sg|}\sum_{\sigma \in \Sg}\sigma^{-1}  \Omega^{\rm{NN}}(\sigma f) \\
    &= \eta \bar{\Omega}( f),
\end{align*}
which concludes the proof. \qed

\end{appendices}

\section*{Acknowledgements}
%
The authors would like to thank Giulio Ortali, Gianluca Di Staso and Yifeng Tian
for useful discussions.
This work has been co-authored by employees of Los Alamos National Laboratory, which 
is operated by Triad National Security, LLC, for the National Nuclear Security 
Administration of U.S. Department of Energy (Contract No. 89233218CNA000001).


\section*{Data Availability Statement}
%
A minimal set of scripts allowing to 
\textit{i)} generate the training dataset 
\textit{ii)} train a neural network and 
\textit{iii)} plug the neural network in a LBM simulation, can be found at 
\url{https://github.com/agabbana/learning_lbm_collision_operator}

\bibliography{biblio}

\end{document}